\documentclass{vgtc}                          

\graphicspath{{figures/}{pictures/}{images/}{./}}

\usepackage{times}

\usepackage{tabu}                      
\usepackage{booktabs}                  
\usepackage{lipsum}                    
\usepackage{mwe}                       
\usepackage{mathptmx}                  
\usepackage{algpseudocode}
\usepackage{amssymb}
\usepackage{enumitem}
\usepackage{multirow}
\usepackage{comment}
\usepackage[caption=false,font=normalsize,labelfont=sf,textfont=sf]{subfig}
\usepackage{amsmath,amssymb}
\usepackage{balance}
\setlist[description]{leftmargin=\parindent,labelindent=0pt,topsep= 0cm, partopsep = 0cm, parsep=0cm,itemsep=0cm}

\newcommand*\samethanks[1][\value{footnote}]{\footnotemark[#1]}

\newtoggle{anonym}
\togglefalse{anonym}

\newcommand{\anonym}[1]{\iftoggle{anonym}{\textbf{ANONYMIZED}}{#1}}

\onlineid{1029}

\vgtccategory{Research}

\vgtcinsertpkg

\title{Investigating Resolution Strategies for Workspace-Occlusion in Augmented Virtuality}

\author{Nico Feld\thanks{e-mail: \{feldn, bimberg, s4mifeld, s4mawoel, weyers\}@uni-trier.de}\\ %
    \scriptsize Trier University %
    \and Pauline Bimberg\samethanks[1]\\ %
    \scriptsize \centering Trier University
\and Michael Feldmann\samethanks[1]\\ %
    \scriptsize \centering Trier University
\and Matthias Wölwer\samethanks[1]\\ %
    \scriptsize \centering Trier University
\and Eike Langbehn\thanks{e-mail: eike.langbehn@haw-hamburg.de}\\ %
    \scriptsize \centering HAW Hamburg
\and Benjamin Weyers\samethanks[1]\\ %
    \scriptsize \centering Trier University
\and Daniel Zielasko\thanks{e-mail: daniel.zielasko@rwth-aachen.de}\\ %
    \scriptsize \centering Trier University
}

\teaser{
  \centering
  \includegraphics[width=\linewidth]{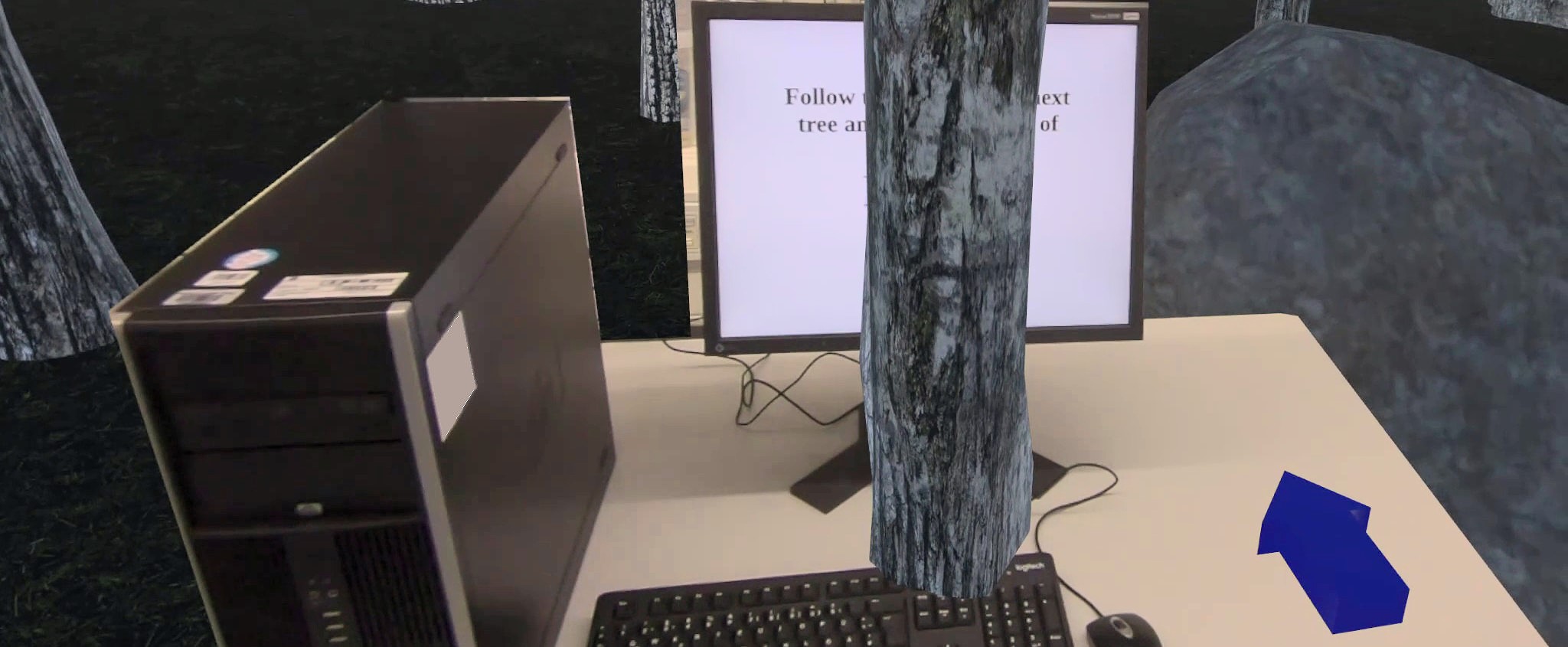}
  \caption{Occurrence of unwanted occlusion when integrating physical content in a virtual environment. Due to a misalignment of the physical and the virtual environment, e.g., caused by virtual locomotion, the user might have difficulties interacting with the PC due to the virtual tree blocking the view. }
  \label{fig:teaser}
}

\abstract{
Augmented Virtuality integrates physical content into virtual environments, but the occlusion of 
physical by virtual content is a challenge.
This unwanted occlusion may disrupt user interactions with physical devices and compromise safety and usability. 
This paper investigates two resolution strategies to address this issue: 
Redirected Walking, which subtly adjusts the user's movement to maintain physical-virtual alignment, and Automatic Teleport Rotation, which realigns the virtual environment during travel.
A user study set in a virtual forest demonstrates that both methods effectively reduce occlusion. 
While in our testbed, Automatic Teleport Rotation achieves higher occlusion resolution, it is suspected to increase cybersickness compared to the less intrusive Redirected Walking approach.
}


\begin{document}

\firstsection{Introduction}

\maketitle
Augmented Virtuality (AV) is a stage within the reality-virtuality continuum of immersive technologies \cite{Milgram1995AugmentedRA} that has seen growing interest in recent years.
Positioned between augmented reality (AR) and virtual reality (VR), it describes the integration of physical content into a predominantly virtual environment.
This physical content can be anything from food~\cite{Budhiraja2015WheresMD,Nakano2022UkemochiAV}, over additional physical devices~\cite{Desai2017AWT,Alaee2018AUS,Shin2022IncorporatingRO,mcgill2015dose} to physical collaborators~\cite{Willich2019YouIM,Pointecker2023VisualMF,Roo2017OneRA}.
Despite its potential, research on AV remained sparse over the years due to technical limitations similar to AR, like pose estimation, combined with additional challenges like the correct incorporation of the physical content inside the virtual environment, for example, the correct occlusion of physical content by virtual content and vice versa \cite{Feld2024EffectsOH}.
However, devices like the Varjo XR-4, the Meta Quest 3, and the Apple Vision Pro are increasingly addressing these limitations through advancements in video-see-through technology.

Still, one of the major challenges of AV is the registration of the physical content in the virtual environment.
While in AR, the virtual content can be manipulated to fit in the physical environment, manipulating the physical content in AV is, in most cases, impossible or requires a complex physical setup, as physical objects cannot be directly controlled without external intervention or specialized hardware.
This issue becomes increasingly challenging when the user can travel in the virtual environment via a virtual locomotion technique, e.g., selection-based teleport.
With virtual locomotion, the user moves only within the virtual environment, causing a misalignment with the physical world.
This misalignment could result in the coincident placement of a physical and a virtual object so that these two objects occlude each other and, thus, make it difficult to interact with them.
For instance, in the case of designing a factory layout while being immersed in a digital twin of this layout, the user's virtual movement may lead to the occlusion of a physical keyboard used for construction on the fly. 
Therefore, an AV application that allows for virtual locomotion has to address unwanted occlusions of physical content through the virtual environment, e.g., by providing strategies to resolve it.
To explore initial approaches for addressing unwanted occlusion, this work examines the following research question:
\textbf{How can unwanted occlusion of a physical object by the virtual environment be solved in an AV setting?}

To investigate this research question, we first explore existing alignment techniques used in related work and how they could be applied to an AV setting to resolve the occlusion of a physical object in Section~\ref{sec:rw}.
To improve external validity, we only consider techniques that require no additional user input and are task agnostic, working without information about the actual task.
Within this scenario, we implement two resolution strategies and investigate their applicability through an exploratory study.
With our work, we contribute to the advancement of AV technology and its practical applications across various use cases.

\section{Related Work}
\label{sec:rw}
This section explores how AV is being used to integrate physical objects into virtual environments to enhance usability and improve user awareness through obstacle and bystander visualization. 
To address the issue of unwanted occlusion, we first look at previous work about passive haptics, as it also relies on precise alignment between physical and virtual objects, similar to the challenges faced in AV. 
We focus on the strategies of redirected walking and teleportation-based alignment, which have been used to maintain consistency in passive haptics and could also help resolve occlusion in AV. 
These studies lay the groundwork for the resolution strategies we present in this work.

\subsection{Augmented Virtuality}

AV bridges the gap between physical and virtual environments by selectively incorporating physical content into virtual environments. 
Its primary applications can be grouped into two areas: (1) integrating physical objects into virtual environments to support interaction and increase usability, and (2) enhancing user awareness of their physical surroundings to improve safety and situational awareness. 
Below, we review key works that demonstrate these capabilities.

\subsubsection{Integration of Physical Content}
Integrating physical objects into virtual environments (cf. \cite{zielasko2017}) is a core application of AV, enabling users to interact with physical tools in a way that supports productivity and increases usability. 
McGill et al.~\cite{mcgill2015dose} explored this potential by incorporating physical objects, such as keyboards, into VR environments. 
Their results showed that selectively blending in physical content improved user performance, reducing typing errors and increasing speed compared to setups in VR. 
This integration also enhanced awareness and interaction, particularly when real-world peripherals or other people were selectively augmented.
Building on this, Chiossi et al.~\cite{Chiossi2024EvaluatingTP} further demonstrated AV's benefits in a user study examining typing performance, focus, and mental effort in AR, AV, and VR. 
They found that AV offered an optimal balance of engagement and cognitive load, resulting in high task performance and fewer distractions than AR or VR. 
These findings reinforce the role of AV in improving workflows by seamlessly combining physical and virtual elements.
Tian et al.~\cite{Tian2019EnhancingAV} further expanded the scope of AV integration by incorporating physical tools and everyday items, such as a notepad, into virtual environments. 
Their augmented physical notepad featured virtual buttons to save and share notes, blending tactile feedback with digital enhancements. 

Other common integrated everyday items are phones \cite{Shin2022IncorporatingRO, Eichhorn2023ShoppingIB,Desai2017AWT,Alaee2018AUS} to integrate a familiar interaction device into the virtual environment, or even food \cite{Budhiraja2015WheresMD,Nakano2022UkemochiAV} to make eating and drinking possible without leaving the virtual environment.
This integration not only supports more natural and intuitive interactions but also accommodates real-world tasks and scenarios. 

\subsubsection{Awareness of Physical Surroundings}                                             
AV enhances users' awareness of their physical surroundings, improving safety and spatial orientation by dynamically integrating real-world information into virtual environments. 
Several studies have explored AV techniques to support obstacle detection and bystander awareness, demonstrating their effectiveness in maintaining presence while ensuring user safety.

To address the challenge of avoiding real-world obstacles in virtual environments, Tian et al.~\cite{Tian2019EnhancingAV} developed a ``hollowed guiding path'' technique that provides real-time visual cues for obstacle awareness. 
Kang and Han~\cite{Kang2020SafeXRAW} introduced ``SafeXR'', a smartphone-based system that detects obstacles and provides optimized alerts, improving task performance and user safety. 
Likewise, Kanamori et al.\cite{Kanamori2018ObstacleAM} showed that using replicas of physical objects enhances spatial perception without reducing presence, and Sousa et al.\cite{Sousa2019SafeWI} extended this idea with ``Augmented Virtual Reality'' techniques that use visual proximity indicators to improve navigation and reduce collisions.
Additionally, Wozniak et al.~\cite{Wozniak2018TowardsUO} demonstrated that unobtrusive placeholder notifications, like virtual trees marking boundaries, can increase spatial awareness and presence.

Beyond obstacle detection, maintaining awareness of bystanders is crucial for collaborative AV setups. 
Willich et al.\cite{Willich2019YouIM} compared avatars, 3D scans, and 2D images for visualizing bystanders and found that while 2D images allow quicker identification, 3D scans offer more precise localization.
Pointecker et al.\cite{Pointecker2023VisualMF} introduced metaphor-based notifications like virtual doors to notify users of approaching bystanders unobtrusively, and Wang et al.~\cite{Wang2022RealityLensAU} presented ``RealityLens'', which lets users selectively blend real-world elements into virtual environments for more natural interactions while maintaining a presence.

These studies demonstrate the potential of AV to combine physical and virtual environments while keeping users aware of their surroundings. 
However, they do not address the challenge of keeping these objects visible and accessible by avoiding unwanted occlusion. 
In this work, we focus on this issue by exploring strategies to maintain the visibility and usability of physical objects in AV spaces.

\subsection{Environment Alignment for Passive Haptics}
One research field addressing the alignment of physical objects in virtual environments is passive haptics.
This approach uses physical objects to simulate the tactile feedback of virtual items~\cite{Hoffman1998PhysicallyTV,Insko2001PassiveHS} (see also substitutional reality \cite{Simeone2015substitutionalReality}).
This technique aligns closely with AV, where physical elements enhance virtual experiences.
The conceptual overlap between AV and passive haptics is evident in works like the ones presented by Palma et al.~\cite{Palma2021AugmentedVU} and Cha et al.~\cite{Cha2022DesignAU}. 
Palma et al.\cite{Palma2021AugmentedVU} used touch-sensitive 3D-printed objects to merge tactile and virtual interactions, while Cha et al.\cite{Cha2022DesignAU} developed a system that dynamically aligns physical tools with their virtual counterparts.
Zielasko et al. \cite{zielasko2019menu} aligned menus with surfaces surrounding the user.
These examples highlight the potential of passive haptics to improve physical-virtual alignment, a challenge that becomes even more significant when users navigate the environment using locomotion techniques~\cite{Thomas2020TowardsPI}.

\subsubsection{Alignment by Redirected Walking}
One approach for aligning physical and virtual environments in passive haptics together with virtual locomotion is to use redirected walking. 
Here, the users' physical movements are slightly amplified or reduced in the virtual environment, so they are unconsciously steered in a desired direction~\cite{Nilsson201815YO}.
Steinicke et al.~\cite{Steinicke2010EstimationOD} showed that by redirecting users on a circle with a radius of $22m$, they have the illusion of being able to walk straight indefinitely.
Redirected walking can consist of up to three different gains that can be applied to the user's movement: rotation gain, translation gain, and curvature gain.
With rotation gains, the user's head rotations, and with translation gains, the user's head translations are amplified or reduced.
When using curvature gains, the user's straight movements are mapped onto a curve rather than a straight line by applying a small virtual rotation on the user when walking.

Although passive haptics require accurate alignment between the physical and virtual environments and redirected walking intentionally breaks this alignment, the two techniques can be combined. 
Kohli et al.\cite{Kohli2005CombiningPH} introduced a concept that merges passive haptics with redirected walking to create a one-to-many mapping between one physical object and multiple virtual objects.
Thomas et al.\cite{Thomas2020TowardsPI} further extended this idea by dynamically mapping virtual objects onto physical ones while avoiding obstacles. 
Williams et al.~\cite{Williams2021ARCAR} investigated aligning both environments via redirected walking so that obstacles in the virtual environment match obstacles in the physical one.
Together, these studies demonstrate that redirected walking can effectively control the alignment of the physical and virtual environments and, thus, may serve as a suitable strategy for resolving unwanted occlusion in AV.

\subsubsection{Alignment during Teleportation}
\label{sec:rw:teleport}
Another technique to align the physical and virtual environment is using the virtual locomotion technique itself.
This is particularly feasible when we focus on target selection based on teleportation (cf. \cite{prithul2021teleportation}), which is a reasonable choice due to its omnipresence in many applications, thanks to its simplicity and ergonomics.
According to Weissker et al., \cite{Weissker2018SpatialUA}, the teleport technique can be separated into four stages.
In the \textit{target specification} stage, the user selects a destination, for example, by pointing.
In the \textit{pre-travel information} stage, the system provides additional information about the teleport, like a preview of the user’s future location and viewing direction, often shown as an avatar; these first two stages can even occur simultaneously.
Then, the actual teleport happens in the \textit{transition} stage, for example, via an instant teleport or during a short fade to black.
Lastly, the system might provide additional information about the teleport in the optional \textit{post-travel feedback} stage, like an arrow to the user's position before the teleport.
Typically, if desired, users must adjust their position and rotation after these stages. 
However, some methods integrate the desired rotation into the target specification phase, either by rotating the controller~\cite{Bozgeyikli2016PointT} or by specifying a point of interest for the user to face after teleporting~\cite{Bimberg2021VirtualRF}.

In AV, where physical objects are integrated into the virtual environment, the \textit{pre-travel information} stage can also display information about these physical contents, such as their positions and extents after the teleport.
Zhang et al.~\cite{Zhang2020VirtualNC} explored three techniques that reposition and reorient the entire workspace in the \textit{target specification} stage, using a cubic frame to show its extent and rotation.
In two of these techniques, the chosen position and rotation are applied to the workspace’s center, one with an avatar preview (\textit{Exo-with-avatar}) and one without (\textit{Exo-without-avatar}).
With the third technique, the user's future position is again previewed as an avatar, but the position and rotation are applied to the user instead of the center of the workspace (\textit{Ego-with-avatar}).
They found that using an avatar significantly improved orientation and target identification time, with users preferring the \textit{Ego-with-avatar} method for better spatial awareness and task efficiency.
Later, Zhang et al.\cite{Zhang2021InTW} proposed a method to support passive haptics alongside teleportation by restricting teleport targets so that the spatial relationship between physical and virtual objects is maintained.
In a similar vein, Wheeler et al.\cite{Wheeler2024ProporientedWR} introduced the ``Prop-Oriented World Rotation'' technique, which realigns the virtual world during teleportation by rotating and translating the user, ensuring that physical objects align with their virtual counterparts. 
The authors demonstrate that this approach simplifies the integration of passive haptics.

In summary, research on passive haptics offers promising approaches for realigning the virtual environment with the physical environment. 
Studies by Kohli et al.~\cite{Kohli2005CombiningPH} and Thomas et al.~\cite{Thomas2020TowardsPI} have shown that redirected walking can effectively align these environments while maintaining the integrity of passive haptic feedback. Additionally, techniques such as the Prop-Oriented World Rotation (POWR) introduced by Wheeler et al.~\cite{Wheeler2024ProporientedWR} realign the virtual world during virtual locomotion.
As both redirected walking and alignment during teleportation have great potential to be able to align the physical and virtual environment for passive haptics, we see the same potential in resolving unwanted occlusion in AV and helping us investigate our research question.
Thus, in the following sections, we propose our implementations for two resolution strategies to solve unwanted occlusion, one based on redirected walking and the other on teleportation.
Then, we propose our task design in Section \ref{sec:task}, which involves physical and virtual locomotion, as well as frequent unwanted occlusions, to investigate both strategies.
The actual study and its results are then described and discussed from Section \ref{sec:eval} onwards.

\section{Resolution Strategies}
\label{sec:strategies}

\begin{figure*}
    \centering
    \includegraphics[width=\linewidth]{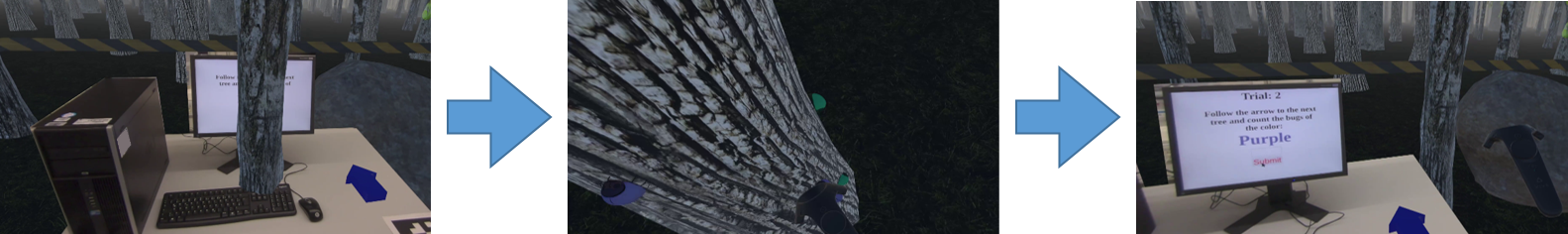}
    \caption{RDW strategy: The desk is initially occluded by a virtual tree (left). During the main task, which involves physical walking around other trees, subtle gains are applied to redirect the user (middle). By the end of the main task, the occlusion is resolved (right).}
    \label{fig:rdw_strategy_flow}
\end{figure*}

\begin{figure*}
    \centering
    \includegraphics[width=\linewidth]{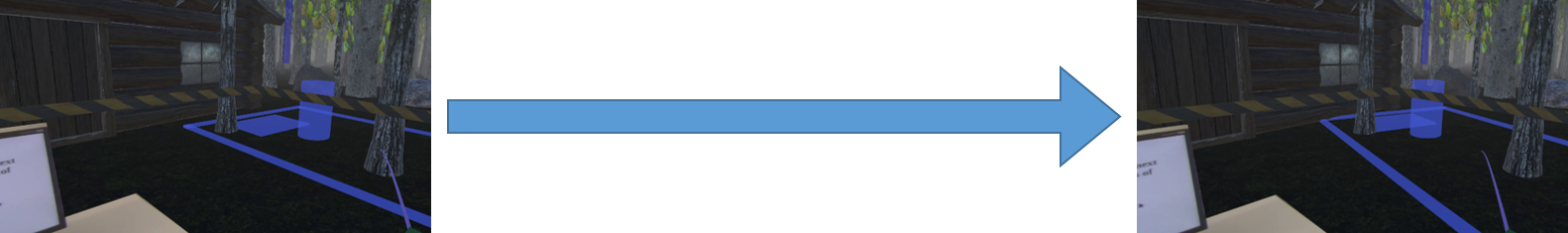}
    \caption{ATR strategy: The left image shows a teleport preview without occlusion, requiring no adjustment. In the right image, the user moved the preview so that the tree would now occlude the desk, prompting the system to automatically rotate the preview upon teleportation.}
    \label{fig:atr_strategy_flow}
\end{figure*}

A strategy to resolve the occlusion of the physical content by the virtual environment in various scenarios must meet certain requirements.
First, the user's intentions are not always known to the system, and therefore, these strategies should be task agnostic.
This means that the strategies can not determine where the user is traveling and which virtual or physical objects might be of interest to the user.
Thus, they can not use this information to resolve the occlusion, e.g., by hiding or moving irrelevant content.
Second, these strategies should not negatively impact task performance.
Thus, they should be automatic instead of being performed by the user, as additional interactions might impact workload, flow, and task performance.
Considering these requirements and based on the related work about environment alignment for passive haptics, utilizing redirected walking and teleportation, we designed two strategies: \textit{Redirected Walking} (RDW) and \textit{Automatic Teleport Rotation} (ATR).

\subsection{Redirected Walking (RDW)}
With RDW, we apply the common redirected walking techniques, rotation gain, translation gain, and curvature gain, to resolve the occlusion during a task.
Redirected Walking techniques usually rely on not being actively noticed by the user \cite{Steinicke2010EstimationOD}.
Gains should, therefore, only be applied when all physical objects are outside of their current field of view.
The specific gains were first based on the investigation of perception thresholds by Steinicke et al. \cite{Steinicke2010EstimationOD} and fine-tuned during testing to find optimal gains for our scenario.
The final gains are $6\%$ for rotation (both amplifying and reducing), $6\%$ for translation (both amplifying and reducing), and $2.6 \frac{\circ}{m}$ for the curvature gains.
All these gains are either lower or equal to the ones proposed by Steinicke et al.~\cite{Steinicke2010EstimationOD}, so they should not be noticeable to the user.
To determine if the user's movement requires amplifying or reduction, the system frequently calculates the closest point (using an algorithm described below) for a physical object to be occlusion-free and applies the correct gains accordingly.

\subsection{Automatic Teleport Rotation (ATR)}
With ATR, we align both environments by virtually rotating the user and the integrated physical objects during the teleportation process so that no occlusions occur between the physical and virtual objects.
This strategy is inspired by the work of Zhang et al.~ \cite{Zhang2021InTW} and Wheeler et al.~\cite{Wheeler2024ProporientedWR}, which used teleportation to align the physical and virtual environments to improve passive haptics.
As the user is not only teleporting themselves but has to consider the placement of physical objects, we use the \textit{Exo-with-avatar} implementation by Zhang et al.~\cite{Zhang2020VirtualNC}, as discussed in Section \ref{sec:rw:teleport}.
When the user selects a target location for the selection-based teleportation in the \textit{target specification} stage that would lead to occlusions, the system again calculates the closest point where the physical object would not be occluded and adjusts the preview accordingly.
To be more in line with the RDW strategy and the included translation gain, this calculation includes not only the rotation around the workspace's origin but also a slight translation of the preview to find an occlusion-free position for the physical object (see below).

In summary, RDW and ATR differ in three aspects:
First, the RDW strategy is applied during real-walking, while the ATR strategy is applied during virtual locomotion.
Second, the RDW strategy is egocentric, as the user is the center of the manipulations, while the ATR strategy is exocentric, as the manipulations are applied on the teleports' preview before the teleportation.
Third, with RDW, the manipulation is applied subtly over time, while with ATR, it is applied instantly and noticeably to the user.

\subsection{Resolution algorithm}
\label{sec:strategies:algorithm}
As both strategies apply a transformation to the user's position and rotation to resolve the occlusion of the physical object, they must determine which transformations to apply and to which degree.
Therefore, both strategies use the same resolution algorithm that calculates the closest occlusion-free point from the physical object's current position or, in the ATR strategy case, the physical object's preview position.
The algorithm is applied every frame while the user specifies the teleport target location (ATR) or is physically walking (RDW).

The algorithm starts with the physical object's initial position and rotation and tries to find the nearest occlusion-free position by incrementally applying rotations and translations, similar to an $A^*$ algorithm.
The maximum angle of the applied rotation is set at $90^\circ$ in either direction, and the maximum distance of the applied translation is set to $1m$.
These constraints were determined during development to minimize the impact on performance in difficult setups while still being able to find an occlusion-free position in most cases.
If no occlusion-free position is found within these constraints, the active resolution strategy is not applied in that frame.
Depending on the strategy, the rotations are calculated around the user's position (RDW) or the workspace-previews center (ATR).

\begin{figure*}
	\centering
	\subfloat[The ATR Strategy]{\includegraphics[width=.49\textwidth]{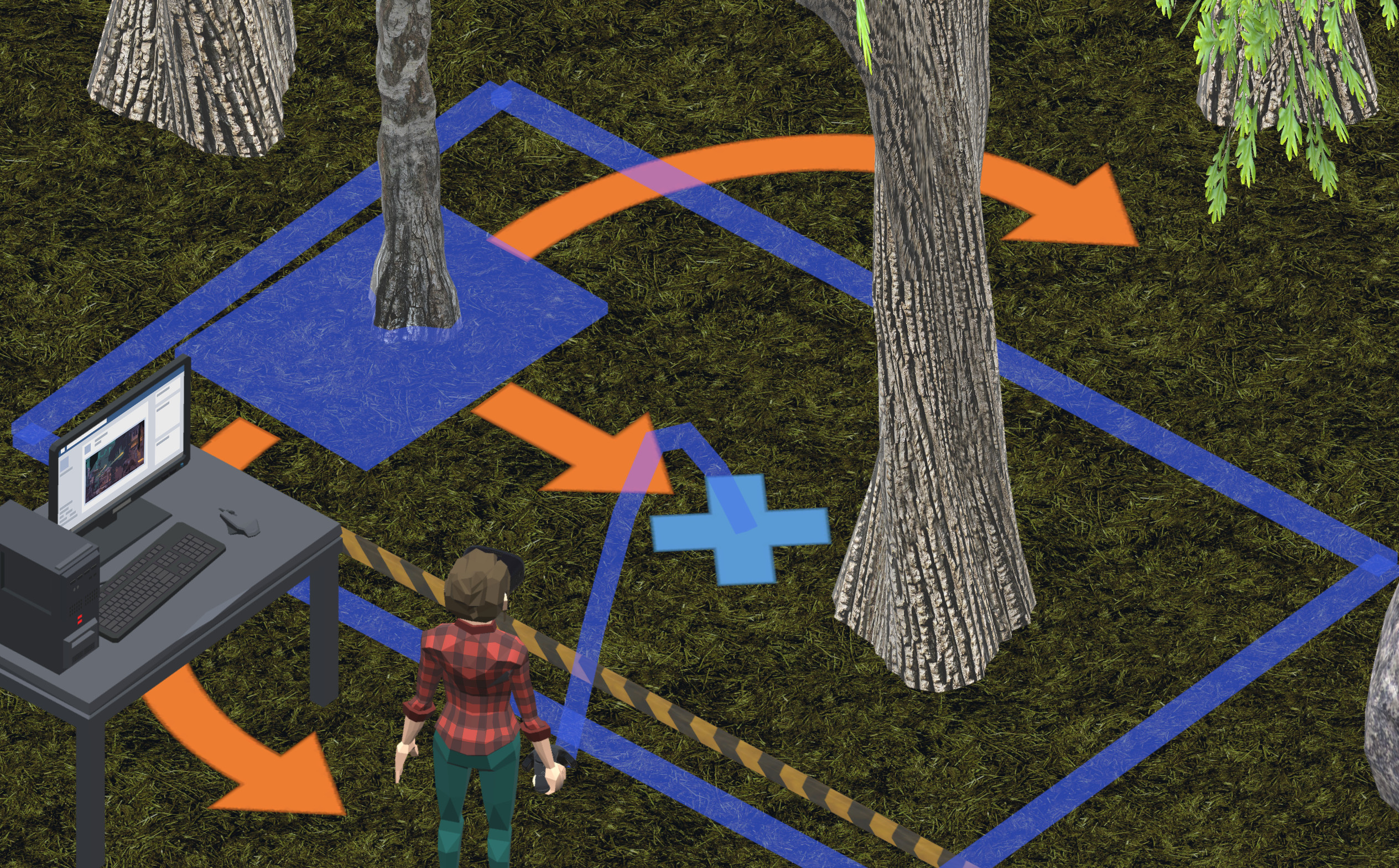}\label{fig:atr_strategy}}
	\hfill
	\subfloat[The RDW Strategy]{\includegraphics[width=.49\textwidth]{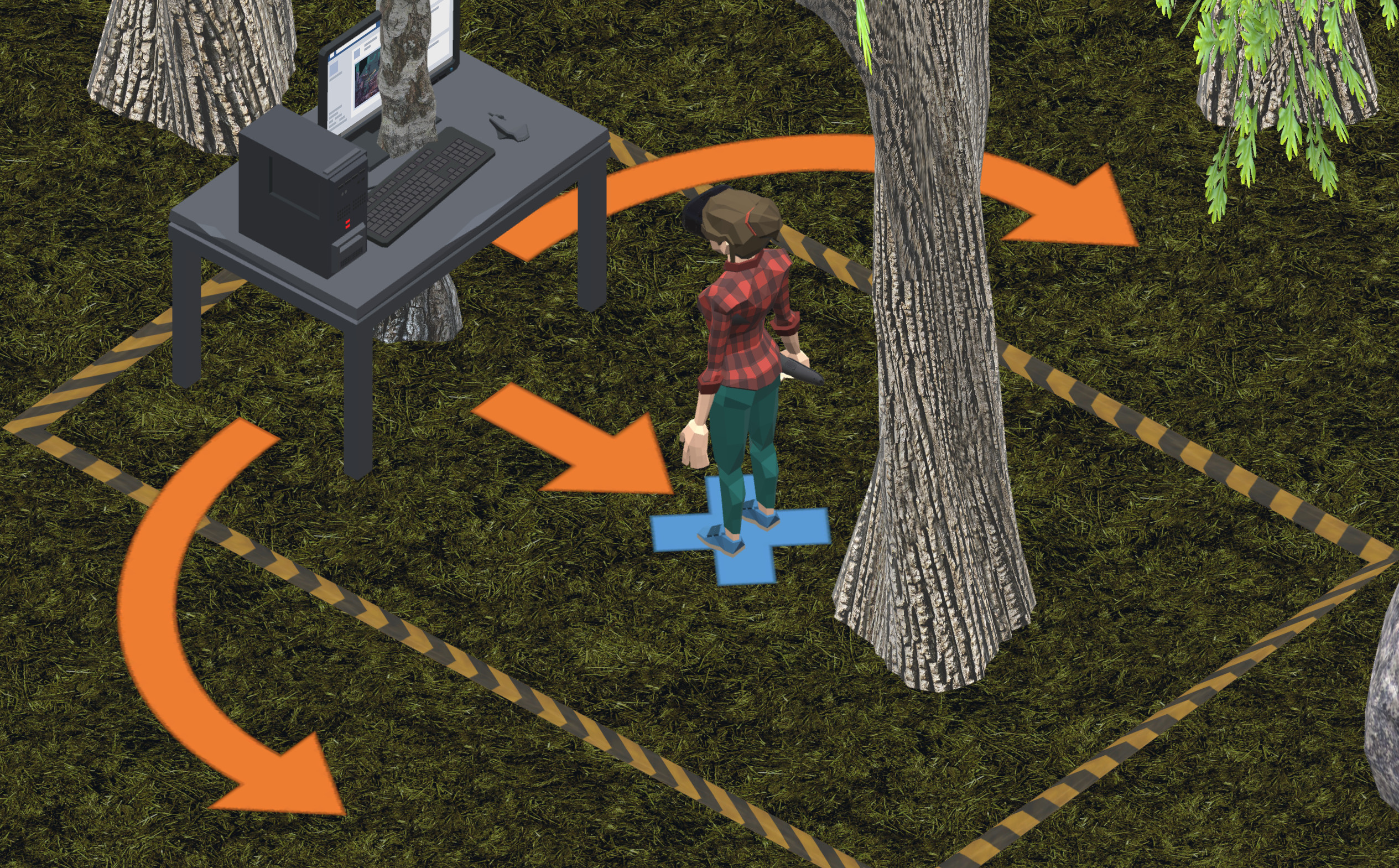}\label{fig:rdw_strategy}}
 \caption{The possible manipulations of the desk to find an occlusion-free position (orange). The translation is always directed towards and the rotations around the origin (blue cross).
 With the ATR strategy (a), the origin is the center of the teleport-preview's position.
    With the RDW strategy (b), the origin is the user's position.}
    \label{fig:strategy}
\end{figure*}

The translations are only applied in the direction of the rotation's origin.
Allowing translations away from the rotation's origin could cause the virtual object to appear between the user and the physical object, causing unwanted obstruction in the workspace.
Orthogonal translation may interfere with position changes resulting from the rotation.
So, as seen in Figure~\ref{fig:strategy}, the desk can, therefore, be rotated clockwise or counterclockwise and can be translated in the direction of the origin.
Finally, the algorithm yields an occlusion-free position that is reachable by applying rotations and translations around the given origin.
It passes that calculated position to the active resolution strategy, which now tries to resolve the occlusion by either applying different gains (RDW) or manipulating the teleport preview (ATR).

\begin{figure}
    \centering
    \begin{align*}
\mathbf{p}' &= \mathbf{p}_{t} + g_t^{\text{dir}} \cdot \frac{\mathbf{p}_{t} - \mathbf{p}_{\text{desk}}}{\|\mathbf{p}_{t} - \mathbf{p}_{\text{desk}}\|} \cdot \|\mathbf{p}_{t-1} - \mathbf{p}_{t}\| \\
\theta' &= \theta_{t} + g_r \cdot \|\theta_{t-1} - \theta_{t}\| + g_c \cdot \|\mathbf{p}_{t-1} - \mathbf{p}_{t}\| \\
\text{where:} & \\
g_t^{\text{dir}} \in \{0.06\} & : \text{Directional translation gain} \\
g_r \in \{-0.06,0.06\} & : \text{Rotation gain} \\
g_c \in \{-2.6,2.6\} & : \text{Curvature gain} \\
\mathbf{p}_{t}, \mathbf{p}_{t-1} & : \text{Current and previous positions of the user} \\
\theta_{t}, \theta_{t-1} & : \text{Current and previous yaw rotations of the user.}
\end{align*}
    \caption{Formular to calculate the updated position ($p'$) and yaw rotation ($\theta'$) with the RDW strategy. The signs of the rotation and curvature gains indicate whether the desk should be rotated clockwise or counterclockwise around the user.}
    \label{fig:rdw_formular}
\end{figure}

Further, for RDW, the rotation direction is locked for the algorithm once the first occlusion-free position is found, so the physical object would not move back into an occluding virtual object when the algorithm might find a closer occlusion-free position in the other rotation direction.
This could be caused by the user's subsequent movements, as the rotation's origin of the algorithm is set to the user's current position.
The rotation direction is unlocked again when the user teleports to a new location.
You can find the full formula to calculate the gains for the RDW strategy in Figure~\ref{fig:rdw_formular}.



\section{Task}
\label{sec:task}
To investigate our strategies, we designed a task that reflects an exemplary AV scenario. 
In cooperation with the forestry office \anonym{Trier}, we developed a task set in a virtual forest, closely resembling common tree assessment tasks.

The virtual forest scenario allows us to strike a balance between a plausible, semi-realistic analytic task and a controlled environment in terms of the number of occlusions by adjusting parameters such as the number of trees and their DBH\footnote{Diameter at Breast Height (DBH): The standard method for measuring the trunk diameter of a standing tree}, thereby isolating potential effects.

In this scenario, the user is tasked with reporting beetle infestations in multiple trees by recording the number of beetles per tree. To facilitate this, the user's physical desk—along with a PC, monitor, mouse, and keyboard—remains visible and usable throughout the task. The monitor displays the current instructions, and when prompted, the user can input infestation data using the physical mouse and keyboard.

In detail, the task consists of three phases: the \textit{travel phase}, the \textit{survey phase}, and the \textit{pointing phase}.
In the \textit{travel phase}, the participant is instructed to travel through the forest to a highlighted tree with the teleport locomotion technique.
When reaching the highlighted tree, the \textit{survey phase} begins.
In the \textit{survey phase}, the participant has to count the bugs of a color that is displayed on the physical monitor, which requires them to physically walk around that tree.
The user then inputs that number on the physical keyboard and submits it with a click on a button using the mouse.
The bugs' colors were selected from a palette designed to be accessible to individuals with color blindness.
Lastly, in the \textit{pointing phase}, the user has to move their controller into a virtual blue arrow on the physical desk and use it to point to the previous tree they examined, which we used for measuring orientation (see Section~\ref{sec:measures:orientation}). 
If they do not remember the position of the previous tree, they are asked to point below the monitor, indicated by a semi-transparent red box below the monitor, to skip the pointing task.
After confirming their selection with a press of the trigger button of the controller, the instructions on the monitor are updated, and the next trial starts with the \textit{travel phase}.

Figure \ref{fig:rdw_strategy_flow} and Figure \ref{fig:atr_strategy_flow} now show how the two strategies are applied during our task to resolve any occlusion that might occur during the task and allow for the interaction with the physical desk setup.
For RDW, the user is unconsciously redirected while recording the beetle infestation and physically walking around the tree.
For ATR, the preview of the teleport gets manipulated to resolve any unwanted occlusion after teleportation when the user navigates to the next tree to survey.

\section{Evaluation}
\label{sec:eval}
We conduct an exploratory within-subject study with two conditions to test the resolution strategies ATR and RDW within the proposed task design.

\subsection{Procedure}
After a short introduction, each participant signs a consent form and starts with one of the two strategies as their first condition, balanced among all participants, and performs 2 test trials and 15 trials per condition.
After finishing all these trials of a condition, the participants have to fill out a questionnaire regarding the current resolution strategy.
The participants can then take a break and, when ready, proceed with the next condition.
While the conditions are balanced, the order of all trials is the same for all participants.
After completing both conditions, the participants are asked to fill out a final questionnaire about their demographics and their assessment of occlusion being a problem in AV and if one or both presented strategies could resolve this issue.
The whole procedure took around 90 minutes and was approved by the ethics council of our institution.

\subsection{Measures}
To investigate our research question on how to solve occlusion in AV, we compare our two resolution strategies regarding multiple measures.
These measures are based on the findings of prior work and the requirements for both resolution strategies, described in Section \ref{sec:strategies}.
We investigate each measure first with a qualitative analysis based on the comments left by the participants and a descriptive analysis.
Based on these results, we investigate each finding further by performing a statistical analysis of objective and subjective data.

\subsubsection{Task Performance}
We first want to investigate how these two strategies might impact task performance.
While we try to minimize the impact of the strategies, e.g., by avoiding additional interactions, we still assume that there is an impact and want to investigate if it differs between the strategies.
To investigate the impact on task performance of both resolution strategies, we measure the average \textit{task time} and the \textit{error rate} of each trial as indicators for efficiency.
As the resolution strategies are applied at different phases of a trial, the \textit{task time} is split into the time the participants spend in the \textit{travel phase} (\textit{task time$_L$}) and in the \textit{survey phase} (\textit{task time$_S$}).
The task time in the \textit{pointing phase} is omitted, as no resolution strategy was present in this phase.

\subsubsection{Orientation}
\label{sec:measures:orientation}
Second, as both strategies involve rotating and translating the user's virtual position by the system, these uncontrolled manipulations might impact their \textbf{orientation}.
This is both true for redirected walking \cite{Nilsson201815YO}, as well as for teleportation, especially with automatic rotations, as discussed by Rahimi et al.~\cite{Rahimi2020SceneTA}.
We measure orientation with multiple error rates of the pointing task and a subjective measure, in line with the evaluations of orientation by Adhikari et al. \cite{Adhikari2022IntegratingCA} and Zielasko et al. \cite{Zielasko2022SystematicDS}.
A detailed description of the exact calculation of each error is provided by Batschelet~\cite{Batschelet1981CircularSI}.
\begin{description}
    \item[Signed Error] which is the mean of the signed pointing errors, which are the angular deviations between the direction the user pointed to and the correct direction to the last tree.
    A value smaller than 0 indicates a general counter-clockwise misperception, while a value greater than 0 indicates a general clockwise misperception of the target position.
    The absolute value of this error indicates the strength of that misperception.
    \item[Absolute Error] which is the mean of the absolute pointing errors, which are the angular deviations between the direction the user points to and the correct direction to the last tree.
    A smaller absolute error indicates a higher accuracy of the pointing task.
    \item[Configuration Error] which is the absolute mean angular deviation of the signed pointing errors.
    A smaller configuration error indicates a higher precision of the pointing task.
    \item[Absolute Ego-Orientation Error] which is the circular mean of the absolute angular deviations.
    A higher value indicates a higher general misconception about the user's self-orientation \cite{Zielasko2022SystematicDS}.
    \item[Subjective Orientation] which is the subjective impact of the resolution strategies on orientation. 
    This is measured through a 5-Point Likert Scale with the question \textit{``How did the tested resolution strategy affect your ability to orient yourself in the forest?''} between \textit{``It greatly worsened my orientation.''}(1) to `\textit{`It greatly improved my orientation.''}(5).
\end{description}

\subsubsection{Agency}

\begin{table}[]
    \centering
    \fontsize{6pt}{7pt}\selectfont
    \begin{tabular}{c|l}
        Construct & Item \\ \hline
        \multirow{ 3}{*}{SoPA} & I was in full control of what I did.\\ 
         & Things I did were subject only to my free will.\\ 
         & I was completely responsible for everything that results from my actions.\\ \hline
        \multirow{ 3}{*}{SoNA} & The consequences of my actions felt like they don’t logically follow my actions.\\ 
         & The outcomes of my actions generally surprised me.\\ 
         & Nothing I did was actually voluntary.\\ 
    \end{tabular}
    \vspace{.2cm}
    \caption{Custom questionnaire to measure the sense of agency. It is a shorter version of the SoAS~\cite{Tapal2017TheSO} and is answered on a 7-Point Likert Scale from \textit{``I strongly disagree''} (1) to \textit{``I strongly agree''} (7)}
    \label{tab:SoAS}
\end{table}

As both resolution strategies are applied by the system and not by the user, we assume they might impact the perceived sense of agency during the task.
The sense of agency describes the subjective perception of the degree to which the user is in control of their body, mind, and environment.
To measure the sense of agency, we use a custom version of the ``Sense of Agency Scale'' (SoAS) by Tapal et al.~\cite{Tapal2017TheSO}.
The SoAS measures the sense of agency in two separate constructs: the sense of positive agency (SoPA), feeling of control, and the sense of negative agency (SoNA), experiencing learned helplessness.
To shorten the questionnaire and fit our scenario, we use only a subset of the SoAS items listed in Table \ref{tab:SoAS}.

\subsubsection{Workload}
Further, we want to compare the effect of both strategies on workload.
While both strategies do not require additional user input and try to minimize the impact on task performance, the difference in subtlety could result in a different impact on workload.
Additionally, Bruder et al.~\cite{Bruder2015CognitiveRD} showed that redirected walking could negatively impact workload.
Further, a high workload could also be an indicator of a lack of orientation, as argued by Zielasko et al.~\cite{Zielasko2024DiscreteVR}.
To measure workload, we use the raw NASA-TLX questionnaire \cite{hart1988development}.

\subsubsection{Cybersickness}
Manipulating the user's rotation and position can cause severe symptoms of cybersickness.
In particular, using redirected walking could introduce additional nausea \cite{Nilsson201815YO}.
To compare the effect of both strategies on cybersickness, as well as to control for possible effects of cybersickness onto other aspects, we measure cybersickness with the \textit{``Fast Motion Sickness Scale''} (FMS) \cite{fms}.

\subsubsection{User Satisfaction}
User Satisfaction is a critical measure that reflects how well the resolution strategies meet the needs and expectations of users during the task. 
To evaluate User Satisfaction, we assess four key measures: preference, usability, user experience, and usefulness.

To capture the participants' preferred resolution strategy, the participants give each strategy a rating on a 10-point Likert scale from 1-10 with the question \textit{``In general, how do you rate the resolution strategy, with 10 being the best score?''}.
To measure the usability and user experience, we use the UMUX-Lite \cite{Lewis2013UMUXLITEWT} and for usefulness, we first asked if the participants generally think that occlusion is a problem in AV with \textit{``Do you generally think it is beneficial to resolve the occlusion of the workstation, e.g., by one of the presented resolution strategies?''} and if answered with \textit{yes}, if RDW, ATR, or neither of both could help resolve an occlusion.

\subsection{Apparatus \& Virtual Environment}
The study was conducted in an empty room with a desk and PC for the experimenter and a desk and PC for the participant.
The desk plus PC for the participant was placed in a $2.5m$ x $4m$ tracking space, in which they were able to move around freely throughout the study.
The boundaries of the tracking space were constantly displayed via a warning tape, similar to the ``Rubber Band'' metaphor used by Wozniak et al.~\cite{Wozniak2018TowardsUO}.
As an AV HMD, we used the Varjo XR-3 as its video-see-through capabilities allow us to display both physical and virtual objects.
The participant's desk was incorporated into the virtual environment via Varjo's masking and marker-tracking capabilities. 
The study task was implemented using Unity 2021.3 and the Varjo XR Unity Plugin 3.6.0.
As the virtual environment, we modeled a ~$250$m x $250$m virtual forest.
We designed that forest based on data about the tree type, height, DBH, and distribution from the forestry office \anonym{Trier} about the ``\anonym{Meulenwald}'' forest.
Still, we increased the DBH of the target trees by 25\%, and we removed the position of the desk from the teleportation preview, even if these changes limit the external validity of our results.
The DBH was increased to make it impossible for the participant to keep the desk always in sight to improve their orientation for the pointing task, which would prevent the RDW from triggering.
This behavior was observed multiple times during testing.
The desk was removed from the teleportation preview to increase the cases of unwanted occlusions and, thus, expose the participant more often to this problem, making the subjective feedback regarding occlusion more proficient.

\subsection{Participants}
A-priori power analysis with a medium effect size of $dz = .5$, due to a lack of expected effect size from prior work, $\alpha = .05$, and $\beta = .2$ results in a sample size of $N = 34$.
Using our university's internal mailing list, we recruited 34 participants with normal or corrected vision as a convenience sample.
Their age range from $19-44$ years with $M=26.47,\ SD=5.12$.
$15 (44\%)$ participants identified as male, $18 (53\%)$ as female, and $1 (3\%)$ as diverse. 
Out of all participants in the study, $24 (71\%)$ did not wear glasses, while $10 (29\%)$ wore glasses.

\section{Results}
\label{sec:results}
Due to the lack of existing theories regarding the resolution strategies, this research adopts an exploratory approach.
We start by examining the general suitability of our study design by investigating the amount of unwanted occlusion and the efficacy of both resolution strategies for resolving it in Section \ref{sec:results:efficacy}.
This helps us to assess the validity of our subsequent exploratory analysis results.
We then examine the qualitative data in Section \ref{sec:results:cybersickness}-\ref{sec:results:orientation}, consisting of comments left by the participants and observations from the experimenter, to identify notable elements or trends that stand out.
Several key topics emerge from this preliminary analysis, which we identify and explore in greater detail by applying a qualitative analysis of the key measures.
However, the complete descriptive results are shown in Table \ref{tab:measures}.

\subsection{Suitability \& Efficacy}
\label{sec:results:efficacy}
To assess the effectiveness of our task design and, thus, the robustness of the results, we first examine whether participants experienced unwanted occlusion and how effectively each strategy resolved it or whether the study design failed to provoke sufficient occlusion for meaningful analysis.
Our data shows that occlusion occurred in $M=91\%,SD=11\%$ of all trials during the last teleport before the \textit{survey phase}. 
On average, the ATR strategy resolved $M=98\%,SD=2\%$ of these cases through applied rotation during teleportation, while the RDW strategy resolved $M=74\%,SD=13\%$ using redirected walking techniques during the \textit{survey phase}.
Although the ATR strategy resolved significantly more cases with $t=10.692,p=<.001$ than RDW, the RDW strategy still managed to resolve the majority of occlusions. 
These results confirm that participants frequently encountered unwanted occlusion and experienced its resolution through both strategies, providing a solid foundation for exploratory analysis of both subjective and objective measures.

\subsection{Cybersickness}
\label{sec:results:cybersickness}
The first pattern we observe is that the majority of comments were regarding cybersickness. 
While for the RDW strategy, one participant states that it was \textit{``much more pleasant''} (P33) than the ATR strategy, multiple participants state that they felt ``dizzy'' (P9, P14, P22, P29, P32) after using this strategy.
For the ATR strategy, there is also one participant stating it was much more \textit{``pleasant and better''} (P8), but multiple others stated that it ``resulted in nausea'' (P4) and that they felt ``dizzy'' (P10, P15, P19, P22, P30, P33) after using the ATR strategy.
Both strategies receive mostly negative comments regarding cybersickness, which is expected, as we did not explicitly ask about it, due to the Complaint Bias\footnote{The Complaint Bias describes the tendency for individuals to be more likely to express dissatisfaction than satisfaction, leading to an overrepresentation of negative feedback in subjective evaluations. \cite{Yechiam2014TheCB}}.
Still, P22 and P33 explicitly state that they felt less dizzy with RDW than ATR, even with RDW being the second condition they experienced.
This is especially interesting, as cybersickness is known for its susceptibility to carry-over effects, which usually leads to a general increase during the exposure time, regardless of the stimuli of the current condition \cite{Zielasko2024CarryOverER}.
The comments of P22 and P33, however, indicate the opposite effect in our study and, thus, that RDW results in less cybersickness than the ATR strategy.
Further, Nilsson et al.\cite{Nilsson201815YO} state that there are indications that redirected walking techniques usually increase cybersickness, depending on the gains used.
On the other hand, as mentioned earlier, Rahimi et al.~\cite{Rahimi2020SceneTA} state that applying a rotation on teleportation results in a slower spatial awareness and, thus, may also impact cybersickness.

To further analyze the effect of the RDW and ATR strategy on cybersickness, we perform a statistical analysis of the FMS score reported by the participants.
To eliminate any potential order effects, we only investigated the FMS score of the first condition performed by each participant and used a t-test for independent samples as suggested by Zielasko et al. \cite{Zielasko2024CarryOverER}.
The independent samples t-test between RDW ($M=1.29, SD=1.69$) and ATR ($M=2.76, SD=2.80$) indicates a marginally non-significant difference, with $t(32)=1.86$, $p=.073$ (Power: $1-\beta = .437$) and an medium effect size of Cohen’s d = $.637$.

Using a more proficient questionnaire, like the SSQ, may allow for a more detailed analysis, especially regarding the symptoms.
However, as cybersickness is not the main part of this analysis, future work requires a more detailed evaluation with a design that acknowledges cybersickness research.
As we only considered the first condition, it was only half the size of N; thus, future work should aim at a larger sample size to achieve a higher power.

\subsection{Preference}
\label{sec:results:preference}
RDW receives positive comments regarding Preference, Usability, and User Experience, highlighting that it was \textit{``quick and fun''} (P27) and \textit{``much easier [than ATR]''} (P11) and only one negative comment, criticizing that the redirected walking would sometimes move the target tree outside the tracking space (P25).
The ATR strategy received multiple positive comments, noting that it was \textit{“quick and pleasant”} (P8, P24), \textit{“good for my purposes”} (P25, 28), and \textit{“better [in general]”} (P8, P22, P28).
On the other hand, the ATR strategy received negative comments, like that the controls feel \textit{``very imprecise''} and \textit{``annoying''} (P4), the \textit{``[...] rotation makes it almost impossible [...] to get where I want to go.''} (P11), that the \textit{``fiddling [...] is really bad''} (P18) and that the strategy is \textit{``very problematic''} (P32) in general.
This polarized feedback regarding the ATR strategy suggests a controversial view on this strategy by users, where some prefer it over the RDW strategy, while others really dislike it.

\begin{figure}
	\centering
	{\includegraphics[width=\linewidth]{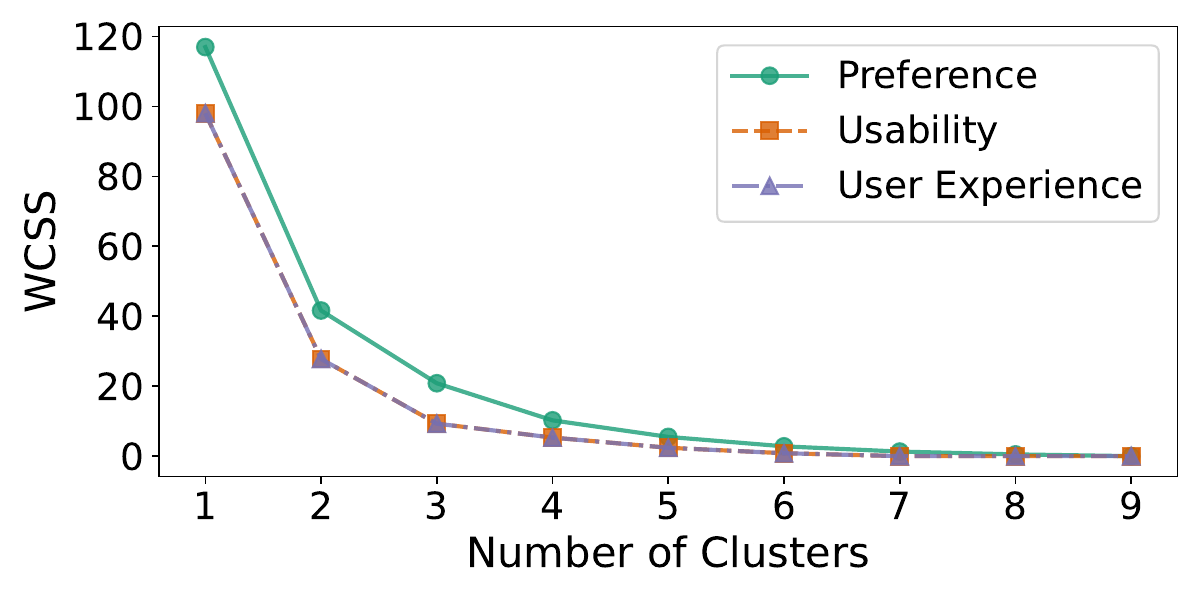}}
 \caption{The sum of squared distances between points and their cluster centers (WCSS) of each number of clusters (k). The k-value, where the WCSS starts to decrease much slower, indicates an optimal clustering. In this case, the plot indicates an optimal clustering with either k=3 or k=4 for Preference and k=3 for Usability and User Experience.}
    \label{fig:elbow}
    \vspace{-.3cm}
\end{figure}

To further investigate this, we use the elbow method\footnote{The Elbow Method helps identify the optimal number of clusters by finding the point where adding clusters stops significantly reducing variance indicated by the within-cluster sum of squares (WCSS)\cite{Thorndike1953WhoBI}.} to investigate optimal clustering in the preference, usability, and user experience data.
As seen in Figure \ref{fig:elbow} the elbow method indicates an optimal clustering with either $k=2$ or $k=3$ for the preference and $k=3$ for the usability and user experience scores.
This supports our assumption that the participants can be divided into distinct groups by the level of user satisfaction and that there is no unified view on this strategy across all participants. 
However, we have no theory as to why ATR is viewed as so controversial in comparison to RDW. 

In terms of the subjective assessment of usefulness regarding both resolution strategies, however, the ATR strategy seems to be perceived as more useful than the RDW strategy:
$17$ of the $34$ participants stated that resolving the occlusion of the workstation is beneficial.
Out of these $17$ participants, $14$ state that ATR is a useful strategy to achieve this, and only $3$ state that it is not useful, indicating the high usefulness of the ATR strategy, emphasized by a Chi-Square test with: $\chi^2 = 7.118, p=.008$.
When assessing the usefulness of the RDW strategy, the distribution is nearly evenly distributed, with $8$ participants finding it useful and $9$ not ($\chi^2 = .059, p =.808$).
None of the $17$ participants stated that neither strategy was useful to resolve the occlusion.
This might indicate that the participants might have controversial views on user satisfaction with the ATR strategy but still thought it is useful in terms of efficacy, which would align with our objective efficacy results discussed in Section~\ref{sec:results:efficacy}.

Comparing both strategies to each other on preference, usability, and user experience reveals no difference between RDW and ATR, as seen in Table \ref{tab:measures}.
Only the user experience rating indicates a possible higher user experience for the RDW strategy with a merely significant p-value of $.055$.

\subsection{Orientation}
\label{sec:results:orientation}
\begin{figure}
    \centering
    \includegraphics[width=\linewidth]{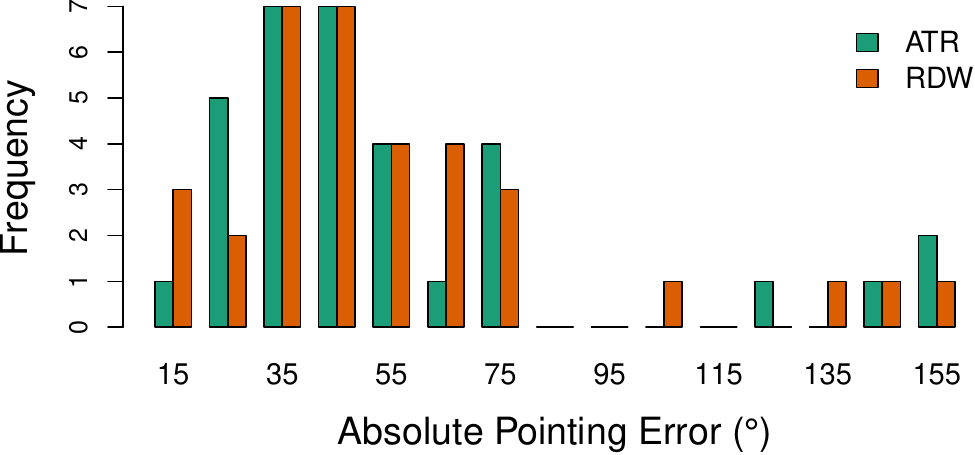}
    \caption{Density of the Absolute Pointing Errors in degrees.}
    \label{fig:orientation_density}
\end{figure}

The feedback from participants indicates significant challenges regarding orientation using the ATR strategy.
Participants like P12 appreciate that the resolution does not occur instantly with the RDW strategy, allowing for some degree of orientation immediately after teleportation. 
Further, many participants express frustration with the automatic positioning in the ATR condition. 
For instance, P25 notes that constant rotation during multiple teleportations leads to confusion, making it difficult to use nearby objects for orientation and often resulting in lost spatial orientation. 
P32 and multiple other participants report feeling disoriented, stating they had ``no idea where I came from'', which underscores a common issue with the method. 
This feedback suggests that the automatic adjustments in ATR reduce the participants' ability to maintain a coherent sense of direction, leading to an overall low spatial orientation during tasks.
To further verify this assumption, we first sanitize the pointing data. As explained in Section \ref{sec:task}, participants have the option to select ``I don't know'' during the pointing task if they can not recall the position of the last tree. 
However, observations during the study indicate that some participants still made random guesses instead of skipping. 
This is supported by Figure \ref{fig:orientation_density}, which shows a cluster of 8 extremely high pointing errors, distinct from the rest.  
Consequently, we exclude the five participants responsible for these errors from further analysis.

The analysis, summarized in Table \ref{tab:measures}, shows no significant differences in Absolute Pointing, Configuration Errors, Absolute Ego-Orientation, or Subjective Orientation between the two strategies. 
This suggests that neither strategy had a distinct impact on pointing accuracy, precision, or self-orientation. 
However, the high means in Absolute Pointing and Configuration Error indicate that participants generally have low accuracy and precision, reflecting poor spatial orientation overall, regardless of the resolution strategy.
For example, Rahimi et al.~\cite{Rahimi2020SceneTA} investigated a teleportation technique with translation and rotation, similar to ATR, resulting in an average pointing error of around 32°.
With a median rating of 3 for both strategies in Subjective Orientation (indicating ``[The strategy] had no effect on my orientation''), we further infer that this low spatial orientation is more likely due to the nature of the task and the virtual environment rather than the strategies themselves.

However, we find a significant difference in Signed Pointing Error between RDW ($M=-9.05, SD=14.81$) and ATR ($M=4.90, SD=24.31$) with $t=-2.890,p=.008$ and an effect size of $d=.546$.
On average, participants using RDW point approximately $9^\circ$ to the left, while using the ATR strategy, they point about $5^\circ$ to the right of the last tree's position.
To investigate whether this difference stems from a rotation bias caused by the resolution strategy, we analyze the Signed Rotation Angle, the average rotation applied by each strategy, and its correlation with the Signed Pointing Error.
The t-test revealed a significant difference between RDW ($M=2.01,SD=3.22$) and ATR ($M=-0.33,SD=2.61$) with $t=2.967,p=.006$ and an effect size of $d=.509$.
Interestingly, while both measures show medium effect sizes, their effects are in opposite directions.
With RDW, participants are rotated on average $2^\circ$ to the right but point on average $9^\circ$ to the left of the target.
For ATR, participants are rotated only on average $0.3^\circ$ to the left, remaining nearly centered, yet pointed on average $5^\circ$ to the right of the target.
Pearson correlations between Signed Rotation Angle and Signed Pointing Error show no significant correlation for RDW with $r(26)=.004, p=.984$, ATR with $r(27)=-.105, p=.587$, or both combined with $r(55)=-.157, p=.244$.
This lack of correlation suggests that both resolution strategies do not directly influence orientation, although RDW appears to exhibit a slight clockwise rotation bias.
However, we have no clear hypothesis to explain the medium effect observed for the Signed Pointing Error or why it is opposite to the effect for the Signed Rotation Angle, and thus, we do not derive any hypothesis at this stage.

\begin{table}[ht]
\resizebox{\linewidth}{!}{%
  \setlength{\tabcolsep}{2pt}%
  \renewcommand{\arraystretch}{0.95}%
  \small
\centering


\begin{tabular}{l|lllll|llllll}
\cline{2-6} \cline{8-12}
 & \multicolumn{5}{l|}{Usability} & \multicolumn{1}{l|}{} & \multicolumn{5}{l|}{User Experience} \\ \cline{2-6} \cline{8-12} 
 & \multicolumn{1}{l|}{$M$} & \multicolumn{1}{l|}{$SD$} & \multicolumn{1}{l|}{$t$} & \multicolumn{1}{l|}{$p$} & $d$ & \multicolumn{1}{l|}{} & \multicolumn{1}{l|}{$M$} & \multicolumn{1}{l|}{$SD$} & \multicolumn{1}{l|}{$t$} & \multicolumn{1}{l|}{$p$} & \multicolumn{1}{l|}{$d$} \\ \cline{1-6} \cline{8-12} 
\multicolumn{1}{|l|}{RDW} & \multicolumn{1}{l|}{$ 5.03 $} & \multicolumn{1}{l|}{$ 1.49 $} & \multicolumn{1}{l|}{\multirow{2}{*}{$ 1.057 $}} & \multicolumn{1}{l|}{\multirow{2}{*}{$ .298 $}} & \multirow{2}{*}{$ .181 $} & \multicolumn{1}{l|}{} & \multicolumn{1}{l|}{$ 5.74 $} & \multicolumn{1}{l|}{$ 1.42 $} & \multicolumn{1}{l|}{\multirow{2}{*}{$ 1.994 $}} & \multicolumn{1}{l|}{\multirow{2}{*}{$ .055 $}} & \multicolumn{1}{l|}{\multirow{2}{*}{$ .342 $}} \\ \cline{1-3} \cline{8-9}
\multicolumn{1}{|l|}{ATR} & \multicolumn{1}{l|}{$ 4.62 $} & \multicolumn{1}{l|}{$ 1.72 $} & \multicolumn{1}{l|}{} & \multicolumn{1}{l|}{} &  & \multicolumn{1}{l|}{} & \multicolumn{1}{l|}{$ 5.00 $} & \multicolumn{1}{l|}{$ 1.69 $} & \multicolumn{1}{l|}{} & \multicolumn{1}{l|}{} & \multicolumn{1}{l|}{} \\ \cline{1-6} \cline{8-12} \noalign{\vskip\doublerulesep\vskip-\arrayrulewidth} \cline{2-6} \cline{8-12}  
 & \multicolumn{5}{l|}{Preference} & \multicolumn{1}{l|}{} & \multicolumn{5}{l|}{NasaTLX} \\ \cline{2-6} \cline{8-12} 
 & \multicolumn{1}{l|}{$M$} & \multicolumn{1}{l|}{$SD$} & \multicolumn{1}{l|}{$t$} & \multicolumn{1}{l|}{$p$} & $d$ & \multicolumn{1}{l|}{} & \multicolumn{1}{l|}{$M$} & \multicolumn{1}{l|}{$SD$} & \multicolumn{1}{l|}{$t$} & \multicolumn{1}{l|}{$p$} & \multicolumn{1}{l|}{$d$} \\ \cline{1-6} \cline{8-12} 
\multicolumn{1}{|l|}{RDW} & \multicolumn{1}{l|}{$ 6.97 $} & \multicolumn{1}{l|}{$ 1.88 $} & \multicolumn{1}{l|}{\multirow{2}{*}{$ .409 $}} & \multicolumn{1}{l|}{\multirow{2}{*}{$ .685 $}} & \multirow{2}{*}{$ .070 $} & \multicolumn{1}{l|}{} & \multicolumn{1}{l|}{$ 38.63 $} & \multicolumn{1}{l|}{$ 16.95 $} & \multicolumn{1}{l|}{\multirow{2}{*}{$ .104 $}} & \multicolumn{1}{l|}{\multirow{2}{*}{$ .917 $}} & \multicolumn{1}{l|}{\multirow{2}{*}{$ .018 $}} \\ \cline{1-3} \cline{8-9}
\multicolumn{1}{|l|}{ATR} & \multicolumn{1}{l|}{$ 6.79 $} & \multicolumn{1}{l|}{$ 2.19 $} & \multicolumn{1}{l|}{} & \multicolumn{1}{l|}{} &  & \multicolumn{1}{l|}{} & \multicolumn{1}{l|}{$ 38.38 $} & \multicolumn{1}{l|}{$ 19.87 $} & \multicolumn{1}{l|}{} & \multicolumn{1}{l|}{} & \multicolumn{1}{l|}{} \\ \cline{1-6} \cline{8-12} \noalign{\vskip\doublerulesep\vskip-\arrayrulewidth} \cline{2-6} \cline{8-12}  
 & \multicolumn{5}{l|}{SoPA} & \multicolumn{1}{l|}{} & \multicolumn{5}{l|}{SoNA} \\ \cline{2-6} \cline{8-12} 
 & \multicolumn{1}{l|}{$M$} & \multicolumn{1}{l|}{$SD$} & \multicolumn{1}{l|}{$t$} & \multicolumn{1}{l|}{$p$} & $d$ & \multicolumn{1}{l|}{} & \multicolumn{1}{l|}{$M$} & \multicolumn{1}{l|}{$SD$} & \multicolumn{1}{l|}{$t$} & \multicolumn{1}{l|}{$p$} & \multicolumn{1}{l|}{$d$} \\ \cline{1-6} \cline{8-12} 
\multicolumn{1}{|l|}{RDW} & \multicolumn{1}{l|}{$ 5.41 $} & \multicolumn{1}{l|}{$ 1.39 $} & \multicolumn{1}{l|}{\multirow{2}{*}{$ .682 $}} & \multicolumn{1}{l|}{\multirow{2}{*}{$ .500 $}} & \multirow{2}{*}{$ .117 $} & \multicolumn{1}{l|}{} & \multicolumn{1}{l|}{$ 5.54 $} & \multicolumn{1}{l|}{$ 1.10 $} & \multicolumn{1}{l|}{\multirow{2}{*}{$ -.419 $}} & \multicolumn{1}{l|}{\multirow{2}{*}{$ .678 $}} & \multicolumn{1}{l|}{\multirow{2}{*}{$ .072 $}} \\ \cline{1-3} \cline{8-9}
\multicolumn{1}{|l|}{ATR} & \multicolumn{1}{l|}{$ 5.21 $} & \multicolumn{1}{l|}{$ 1.50 $} & \multicolumn{1}{l|}{} & \multicolumn{1}{l|}{} &  & \multicolumn{1}{l|}{} & \multicolumn{1}{l|}{$ 5.62 $} & \multicolumn{1}{l|}{$ 0.96 $} & \multicolumn{1}{l|}{} & \multicolumn{1}{l|}{} & \multicolumn{1}{l|}{} \\ \cline{1-6} \cline{8-12} \noalign{\vskip\doublerulesep\vskip-\arrayrulewidth} \cline{2-6} \cline{8-12}  
 & \multicolumn{5}{l|}{FMS} & \multicolumn{1}{l|}{} & \multicolumn{5}{l|}{FMS (First Only), N = 17} \\ \cline{2-6} \cline{8-12} 
 & \multicolumn{1}{l|}{$M$} & \multicolumn{1}{l|}{$SD$} & \multicolumn{1}{l|}{$t$} & \multicolumn{1}{l|}{$p$} & $d$ & \multicolumn{1}{l|}{} & \multicolumn{1}{l|}{$M$} & \multicolumn{1}{l|}{$SD$} & \multicolumn{1}{l|}{$t$} & \multicolumn{1}{l|}{$p$} & \multicolumn{1}{l|}{$d$} \\ \cline{1-6} \cline{8-12} 
\multicolumn{1}{|l|}{RDW} & \multicolumn{1}{l|}{$ 2.76 $} & \multicolumn{1}{l|}{$ 2.50 $} & \multicolumn{1}{l|}{\multirow{2}{*}{$ .378 $}} & \multicolumn{1}{l|}{\multirow{2}{*}{$ .708 $}} & \multirow{2}{*}{$ .065 $} & \multicolumn{1}{l|}{} & \multicolumn{1}{l|}{$ 1.29 $} & \multicolumn{1}{l|}{$ 1.69 $} & \multicolumn{1}{l|}{\multirow{2}{*}{$ -1.857 $}} & \multicolumn{1}{l|}{\multirow{2}{*}{$ .075 $}} & \multicolumn{1}{l|}{\multirow{2}{*}{$ .412 $}} \\ \cline{1-3} \cline{8-9}
\multicolumn{1}{|l|}{ATR} & \multicolumn{1}{l|}{$ 2.59 $} & \multicolumn{1}{l|}{$ 2.93 $} & \multicolumn{1}{l|}{} & \multicolumn{1}{l|}{} &  & \multicolumn{1}{l|}{} & \multicolumn{1}{l|}{$ 2.76 $} & \multicolumn{1}{l|}{$ 2.80 $} & \multicolumn{1}{l|}{} & \multicolumn{1}{l|}{} & \multicolumn{1}{l|}{} \\ \cline{1-6} \cline{8-12} \noalign{\vskip\doublerulesep\vskip-\arrayrulewidth} \cline{2-6} \cline{8-12}  
 & \multicolumn{5}{l|}{Error Rate} & \multicolumn{1}{l|}{} & \multicolumn{5}{l|}{Absolute Errors} \\ \cline{2-6} \cline{8-12} 
 & \multicolumn{1}{l|}{$M$} & \multicolumn{1}{l|}{$SD$} & \multicolumn{1}{l|}{$t$} & \multicolumn{1}{l|}{$p$} & $d$ & \multicolumn{1}{l|}{} & \multicolumn{1}{l|}{$M$} & \multicolumn{1}{l|}{$SD$} & \multicolumn{1}{l|}{$t$} & \multicolumn{1}{l|}{$p$} & \multicolumn{1}{l|}{$d$} \\ \cline{1-6} \cline{8-12} 
\multicolumn{1}{|l|}{RDW} & \multicolumn{1}{l|}{$ .33 $} & \multicolumn{1}{l|}{$ .16 $} & \multicolumn{1}{l|}{\multirow{2}{*}{$ -.695 $}} & \multicolumn{1}{l|}{\multirow{2}{*}{$ .492 $}} & \multirow{2}{*}{$ .119 $} & \multicolumn{1}{l|}{} & \multicolumn{1}{l|}{$ .49 $} & \multicolumn{1}{l|}{$ .23 $} & \multicolumn{1}{l|}{\multirow{2}{*}{$ -1.029 $}} & \multicolumn{1}{l|}{\multirow{2}{*}{$ .311 $}} & \multicolumn{1}{l|}{\multirow{2}{*}{$ .177 $}} \\ \cline{1-3} \cline{8-9}
\multicolumn{1}{|l|}{ATR} & \multicolumn{1}{l|}{$ .35 $} & \multicolumn{1}{l|}{$ .15 $} & \multicolumn{1}{l|}{} & \multicolumn{1}{l|}{} &  & \multicolumn{1}{l|}{} & \multicolumn{1}{l|}{$ .53 $} & \multicolumn{1}{l|}{$ .23 $} & \multicolumn{1}{l|}{} & \multicolumn{1}{l|}{} & \multicolumn{1}{l|}{} \\ \cline{1-6} \cline{8-12} \noalign{\vskip\doublerulesep\vskip-\arrayrulewidth} \cline{2-6} \cline{8-12}  
 & \multicolumn{5}{l|}{Travel Time (s)} & \multicolumn{1}{l|}{} & \multicolumn{5}{l|}{Survey Time (s)} \\ \cline{2-6} \cline{8-12} 
 & \multicolumn{1}{l|}{$M$} & \multicolumn{1}{l|}{$SD$} & \multicolumn{1}{l|}{$t$} & \multicolumn{1}{l|}{$p$} & $d$ & \multicolumn{1}{l|}{} & \multicolumn{1}{l|}{$M$} & \multicolumn{1}{l|}{$SD$} & \multicolumn{1}{l|}{$t$} & \multicolumn{1}{l|}{$p$} & \multicolumn{1}{l|}{$d$} \\ \cline{1-6} \cline{8-12} 
\multicolumn{1}{|l|}{RDW} & \multicolumn{1}{l|}{$ 13.52 $} & \multicolumn{1}{l|}{$ 6.03 $} & \multicolumn{1}{l|}{\multirow{2}{*}{$ .937 $}} & \multicolumn{1}{l|}{\multirow{2}{*}{$ .356 $}} & \multirow{2}{*}{$ .161 $} & \multicolumn{1}{l|}{} & \multicolumn{1}{l|}{$ 38.35 $} & \multicolumn{1}{l|}{$ 13.32 $} & \multicolumn{1}{l|}{\multirow{2}{*}{$ -.478 $}} & \multicolumn{1}{l|}{\multirow{2}{*}{$ .636 $}} & \multicolumn{1}{l|}{\multirow{2}{*}{$ -.082 $}} \\ \cline{1-3} \cline{8-9}
\multicolumn{1}{|l|}{ATR} & \multicolumn{1}{l|}{$ 12.52 $} & \multicolumn{1}{l|}{$ 5.58 $} & \multicolumn{1}{l|}{} & \multicolumn{1}{l|}{} &  & \multicolumn{1}{l|}{} & \multicolumn{1}{l|}{$ 38.96 $} & \multicolumn{1}{l|}{$ 15.19 $} & \multicolumn{1}{l|}{} & \multicolumn{1}{l|}{} & \multicolumn{1}{l|}{} \\ \cline{1-6} \cline{8-12} \noalign{\vskip\doublerulesep\vskip-\arrayrulewidth} \cline{2-6} \cline{8-12}  
 & \multicolumn{5}{l|}{Task Time (Travel Time + Survey Time)} & \multicolumn{1}{l|}{} & \multicolumn{5}{l|}{Subjective Orientation} \\ \cline{2-6} \cline{8-12} 
 & \multicolumn{1}{l|}{$M$} & \multicolumn{1}{l|}{$SD$} & \multicolumn{1}{l|}{$t$} & \multicolumn{1}{l|}{$p$} & $d$ & \multicolumn{1}{l|}{} & \multicolumn{1}{l|}{$Mdn$} & \multicolumn{1}{l|}{$IQR$} & \multicolumn{1}{l|}{$Z$} & \multicolumn{1}{l|}{$p$} & \multicolumn{1}{l|}{$r$} \\ \cline{1-6} \cline{8-12} 
\multicolumn{1}{|l|}{RDW} & \multicolumn{1}{l|}{$ 51.86 $} & \multicolumn{1}{l|}{$ 16.88 $} & \multicolumn{1}{l|}{\multirow{2}{*}{$ .200 $}} & \multicolumn{1}{l|}{\multirow{2}{*}{$ .843 $}} & \multirow{2}{*}{$ .034 $} & \multicolumn{1}{l|}{} & \multicolumn{1}{l|}{$ 3.00 $} & \multicolumn{1}{l|}{$ 1.00 $} & \multicolumn{1}{l|}{\multirow{2}{*}{$ -1.744 $}} & \multicolumn{1}{l|}{\multirow{2}{*}{$ .184 $}} & \multicolumn{1}{l|}{\multirow{2}{*}{$ .299 $}} \\ \cline{1-3} \cline{8-9}
\multicolumn{1}{|l|}{ATR} & \multicolumn{1}{l|}{$ 51.48 $} & \multicolumn{1}{l|}{$ 17.35 $} & \multicolumn{1}{l|}{} & \multicolumn{1}{l|}{} &  & \multicolumn{1}{l|}{} & \multicolumn{1}{l|}{$ 3.00 $} & \multicolumn{1}{l|}{$ 2.00 $} & \multicolumn{1}{l|}{} & \multicolumn{1}{l|}{} & \multicolumn{1}{l|}{} \\ \cline{1-6} \cline{8-12} \noalign{\vskip\doublerulesep\vskip-\arrayrulewidth} \cline{2-6} \cline{8-12}  
 & \multicolumn{5}{l|}{Absolute Pointing Error} & \multicolumn{1}{l|}{} & \multicolumn{5}{l|}{Signed Pointing Error} \\ \cline{2-6} \cline{8-12} 
 & \multicolumn{1}{l|}{$M$} & \multicolumn{1}{l|}{$SD$} & \multicolumn{1}{l|}{$t$} & \multicolumn{1}{l|}{$p$} & $d$ & \multicolumn{1}{l|}{} & \multicolumn{1}{l|}{$M$} & \multicolumn{1}{l|}{$SD$} & \multicolumn{1}{l|}{$t$} & \multicolumn{1}{l|}{$p$} & \multicolumn{1}{l|}{$d$} \\ \cline{1-6} \cline{8-12} 
\multicolumn{1}{|l|}{RDW} & \multicolumn{1}{l|}{$ 43.41 $} & \multicolumn{1}{l|}{$ 15.33 $} & \multicolumn{1}{l|}{\multirow{2}{*}{$ -.575 $}} & \multicolumn{1}{l|}{\multirow{2}{*}{$ .570 $}} & \multirow{2}{*}{$ .109 $} & \multicolumn{1}{l|}{} & \multicolumn{1}{l|}{$-9.95$} & \multicolumn{1}{l|}{$ 14.81 $} & \multicolumn{1}{l|}{\multirow{2}{*}{$ -2.890 $}} & \multicolumn{1}{l|}{\multirow{2}{*}{$ .008 $}} & \multicolumn{1}{l|}{\multirow{2}{*}{$ -0.546 $}} \\ \cline{1-3} \cline{8-9}
\multicolumn{1}{|l|}{ATR} & \multicolumn{1}{l|}{$ 45.27 $} & \multicolumn{1}{l|}{$ 17.44 $} & \multicolumn{1}{l|}{} & \multicolumn{1}{l|}{} &  & \multicolumn{1}{l|}{} & \multicolumn{1}{l|}{$ 4.90 $} & \multicolumn{1}{l|}{$ 24.31 $} & \multicolumn{1}{l|}{} & \multicolumn{1}{l|}{} & \multicolumn{1}{l|}{} \\ \cline{1-6} \cline{8-12} \noalign{\vskip\doublerulesep\vskip-\arrayrulewidth} \cline{2-6} \cline{8-12}  
 & \multicolumn{5}{l|}{Absolute Ego-Orientation Error} & \multicolumn{1}{l|}{} & \multicolumn{5}{l|}{Configuration Error} \\ \cline{2-6} \cline{8-12} 
 & \multicolumn{1}{l|}{$M$} & \multicolumn{1}{l|}{$SD$} & \multicolumn{1}{l|}{$t$} & \multicolumn{1}{l|}{$p$} & $d$ & \multicolumn{1}{l|}{} & \multicolumn{1}{l|}{$M$} & \multicolumn{1}{l|}{$SD$} & \multicolumn{1}{l|}{$t$} & \multicolumn{1}{l|}{$p$} & \multicolumn{1}{l|}{$d$} \\ \cline{1-6} \cline{8-12} 
\multicolumn{1}{|l|}{RDW} & \multicolumn{1}{l|}{$ 14.24 $} & \multicolumn{1}{l|}{$ 10.62 $} & \multicolumn{1}{l|}{\multirow{2}{*}{$ -.175 $}} & \multicolumn{1}{l|}{\multirow{2}{*}{$ .862 $}} & \multirow{2}{*}{$ .033 $} & \multicolumn{1}{l|}{} & \multicolumn{1}{l|}{$ 45.77$} & \multicolumn{1}{l|}{$ 14.58 $} & \multicolumn{1}{l|}{\multirow{2}{*}{$ -.545 $}} & \multicolumn{1}{l|}{\multirow{2}{*}{$ .590 $}} & \multicolumn{1}{l|}{\multirow{2}{*}{$ .103 $}} \\ \cline{1-3} \cline{8-9}
\multicolumn{1}{|l|}{ATR} & \multicolumn{1}{l|}{$ 14.84 $} & \multicolumn{1}{l|}{$ 15.61 $} & \multicolumn{1}{l|}{} & \multicolumn{1}{l|}{} &  & \multicolumn{1}{l|}{} & \multicolumn{1}{l|}{$ 47.29 $} & \multicolumn{1}{l|}{$ 14.33 $} & \multicolumn{1}{l|}{} & \multicolumn{1}{l|}{} & \multicolumn{1}{l|}{} \\ \cline{1-6} \cline{8-12} \noalign{\vskip\doublerulesep\vskip-\arrayrulewidth} \cline{2-6} \cline{8-12}  
 & \multicolumn{5}{l|}{Absolute Rotation Angle} & \multicolumn{1}{l|}{} & \multicolumn{5}{l|}{Signed Rotation Angle} \\ \cline{2-6} \cline{8-12} 
 & \multicolumn{1}{l|}{$M$} & \multicolumn{1}{l|}{$SD$} & \multicolumn{1}{l|}{$t$} & \multicolumn{1}{l|}{$p$} & $d$ & \multicolumn{1}{l|}{} & \multicolumn{1}{l|}{$M$} & \multicolumn{1}{l|}{$SD$} & \multicolumn{1}{l|}{$t$} & \multicolumn{1}{l|}{$p$} & \multicolumn{1}{l|}{$d$} \\ \cline{1-6} \cline{8-12} 
\multicolumn{1}{|l|}{RDW} & \multicolumn{1}{l|}{$ 8.90 $} & \multicolumn{1}{l|}{$ 3.16 $} & \multicolumn{1}{l|}{\multirow{2}{*}{$ -3.744 $}} & \multicolumn{1}{l|}{\multirow{2}{*}{$ .001 $}} & \multirow{2}{*}{$ -0.642 $} & \multicolumn{1}{l|}{} & \multicolumn{1}{l|}{$ 2.01 $} & \multicolumn{1}{l|}{$ 3.22 $} & \multicolumn{1}{l|}{\multirow{2}{*}{$ 2.967 $}} & \multicolumn{1}{l|}{\multirow{2}{*}{$ .006 $}} & \multicolumn{1}{l|}{\multirow{2}{*}{$ .509 $}} \\ \cline{1-3} \cline{8-9}
\multicolumn{1}{|l|}{ATR} & \multicolumn{1}{l|}{$ 11.71 $} & \multicolumn{1}{l|}{$ 2.74 $} & \multicolumn{1}{l|}{} & \multicolumn{1}{l|}{} &  & \multicolumn{1}{l|}{} & \multicolumn{1}{l|}{$ -0.33 $} & \multicolumn{1}{l|}{$ 2.61 $} & \multicolumn{1}{l|}{} & \multicolumn{1}{l|}{} & \multicolumn{1}{l|}{} \\ \cline{1-6} \cline{8-12} \noalign{\vskip\doublerulesep\vskip-\arrayrulewidth} \cline{2-6} 
 & \multicolumn{5}{l|}{Occlusions resolved} &  & \multicolumn{5}{l}{} \\ \cline{2-6}
 & \multicolumn{1}{l|}{$M$} & \multicolumn{1}{l|}{$SD$} & \multicolumn{1}{l|}{$t$} & \multicolumn{1}{l|}{$p$} & $d$ &  &  &  &  &  &  \\ \cline{1-6}
\multicolumn{1}{|l|}{RDW} & \multicolumn{1}{l|}{$ .74 $} & \multicolumn{1}{l|}{$ .13 $} & \multicolumn{1}{l|}{\multirow{2}{*}{$ -10.692$}} & \multicolumn{1}{l|}{\multirow{2}{*}{$ <.001 $}} & \multirow{2}{*}{$ 1.834 $} &  &  &  & \multirow{2}{*}{} & \multirow{2}{*}{} & \multirow{2}{*}{} \\ \cline{1-3}
\multicolumn{1}{|l|}{ATR} & \multicolumn{1}{l|}{$ .98 $} & \multicolumn{1}{l|}{$ .02 $} & \multicolumn{1}{l|}{} & \multicolumn{1}{l|}{} &  &  &  &  &  &  &  \\ \cline{1-6}
\end{tabular}
}
\vspace{.1cm}
\caption{Summary of Descriptive Statistics and Statistical Test Results\label{tab:measures}}
\end{table}

\section{Discussion}
The present study investigated two resolution strategies for addressing unwanted occlusion in AV environments: 
Redirected Walking (RDW) and Automatic Teleport Rotation (ATR). 
By examining aspects such as suitability, efficacy, cybersickness, user satisfaction, and orientation, we aim to evaluate their respective impacts on user experience and effectiveness in resolving occlusion. 
Based on the qualitative feedback, we mainly focus on cybersickness, user satisfaction, and orientation in our exploratory analysis. 
Both the qualitative and quantitative analyses reveal no substantial indications of effects on task performance, agency, or workload.

Our study demonstrates that the task design successfully induces unwanted occlusion in both experimental conditions, validating the effectiveness of the study stimulus. 
Both RDW and ATR effectively resolve occlusion cases, with ATR achieving a significantly higher resolution rate.
Therefore, ATR is suitable for scenarios where the resolution of occlusion is a high priority.
While RDW’s subtler approach resolves fewer resolutions overall, it still addresses most cases, highlighting its viability in scenarios where gradual and less perceptible interventions are preferable.

Based on user reports, both resolution strategies elicit some degree of cybersickness, with ATR associated with marginally higher levels than RDW. 
These observations do not align with our expectations, as teleportation usually does not induce cybersickness \cite{Langbehn2018EvaluationOL}, while redirected walking is often associated with higher levels of cybersickness \cite{Nilsson201815YO}. 
One possible explanation could be the applied rotational adjustments during teleportation in the ATR condition.
According to Rahimi et al., ~\cite{Rahimi2020SceneTA}, viewpoint changes, like teleportation, that not only include a position but also a rotation change might lead to a higher cybersickness.
Further, manipulation in the ATR condition is potentially applied every time the user teleports, while manipulation is only applied once in the RDW condition.
Therefore, the user is exposed to manipulation in the ATR condition more often than in the RDW condition.
Additionally, the selected gains for redirected walking seem to be small enough to avoid contributing significantly to cybersickness.
The average framerates were $89.95$ FPS for ATR and $89.99$ FPS for RDW, which indicates a stable performance of the application for both strategies.
However, as mentioned in Section \ref{sec:results:cybersickness}, our method to measure cybersickness is limited to the FMS questionnaire.
As argued by Slater et al.~\cite{Slater2004HowCW}, it can be hard to quantify how other factors, like usability, user experience, or mental load, affected the participant's answers in a way that makes it hard to extract only the effect of cybersickness, as the other factors might have superimposed cybersickness.

User feedback on satisfaction is mixed. 
While qualitative results indicate balanced opinions for RDW, feedback for ATR is heavily polarized, with some participants favoring its efficiency and clarity in resolving occlusion, while others find it intrusive. 
This polarization is further reflected in the clustering of qualitative measures such as preference, usability, and user experience. 
However, ATR is rated with high usefulness, indicating that participants perceive it as controversial in terms of user satisfaction but generally positive in terms of efficacy.

Contrary to previous findings, such as those of Langbehn et al.~\cite{Langbehn2018EvaluationOL}, which identified significant orientation differences between redirected walking and teleportation techniques, our analysis does not reveal significant orientation effects between RDW and ATR. 
One possible explanation of the different results lies in the consistent use of teleportation as the main locomotion technique across all conditions in our study, with mostly only rotational adjustments during the \textit{travel phase} by ATR or during the \textit{survey phase} by RDW.
Further, we assume that the layout and design of our environment have a negative impact on the general orientation of the participants.
The fidelity of visual cues is very limited, as there is only a small variety of trees and rocks, and no real landmarks are present in the whole environment.
So, in summary, we assume that participants mainly lose orientation due to the use of teleportation as the main locomotion technique and the lack of visual cues in our virtual environment.
Therefore, the possible effects of the resolution strategies are too small to significantly affect the already impaired spatial orientation.
Still, for RDW, we observe a general tendency of participants to point more to the left of the target while being rotated more to the right by the strategy.
Currently, we have no theoretical explanation for these opposing effects, suggesting an area for future exploration.

\section{Limitations \& Future Work}
This study provides valuable insights into resolution strategies for addressing unwanted occlusion in AV environments. 
However, several limitations should be noted to contextualize the findings.

The experimental setup incorporates only one static physical object into the virtual environment, limiting the generalizability of the findings. 
Scenarios involving multiple or dynamic physical objects could introduce complex interactions, potentially affecting the applicability of the resolution strategies.
Additionally, the design and layout of the virtual environment likely impair participants' spatial orientation. 
The repetitive nature of visual elements, such as trees and rocks, and the lack of distinct landmarks may overshadow the effects of the resolution strategies. 
Future studies should include environments with more visual cues to better assess these impacts.

Further, in our study, we focused on resolution strategies that are system-controlled and task-agnostic. 
This approach was chosen to reduce any negative impact on task performance and usability while also making the strategies more broadly applicable across different scenarios. 
However, future research could explore strategies that are tailored to specific tasks or controlled by users. 
These strategies might offer additional insights into how unwanted occlusion can be resolved more effectively, especially in situations where more context about the task or user intentions is available.

These limitations underscore the need for further research with more complex setups, dynamic physical objects, and alternative strategies to comprehensively evaluate the proposed ones.

\section{Conclusion}
This study explores two resolution strategies, Redirected Walking (RDW) and Automatic Teleport Rotation (ATR), to address unwanted occlusion in AV environments. 
Both strategies effectively resolve occlusions, with ATR achieving higher resolution rates and RDW offering a more gradual and less perceptible intervention.
Cybersickness might be slightly higher with ATR than with RDW. 
User satisfaction was mixed, while RDW received balanced feedback and ATR polarized opinions.
Orientation results show no significant differences between strategies, though opposing trends in pointing accuracy and rotation alignment emerged, which remain unexplained. 
The lack of distinct landmarks in the environment likely impairs spatial orientation, overshadowing the strategies' effects.
Our study shows that both RDW and ATR effectively solve the problem of unwanted occlusion in AV. 
Each approach has its strengths and weaknesses.
ATR resolves occlusions more reliably, but it may also cause more cybersickness and is more controversial in preference. 
RDW offers a gentler, less noticeable solution, though it might not always clear occlusions as effectively.
Ultimately, the best method depends on the specific situation and the user’s preferences. 
In practice, we recommend using ATR if resolving occlusion is crucial or physical walking might be limited.
We recommend RDW if unwanted occlusion could occur from time to time, physical walking is a main part of the task, and high user satisfaction should be achieved.

\acknowledgments{
We thank the forestry department of the City of Trier, especially Gundolf Bartmann and Alena Wehr, for their help on the task design and for providing the essential data on the Meulenwald forest.
}

\bibliographystyle{abbrv-doi}

\balance
\bibliography{template}

\begin{thebibliography}{10}

\bibitem{Adhikari2022IntegratingCA}
A.~Adhikari, D.~Zielasko, I.~Aguilar, A.~Bretin, E.~Kruijff, M.~von~der Heyde, and B.~Riecke.
\newblock Integrating continuous and teleporting vr locomotion into a seamless ‘hyperjump’ paradigm.
\newblock {\em IEEE Transactions on Visualization and Computer Graphics}, 29:5265--5281, 2022. doi: {{%
10\hspace{.1pt}\discretionary{.}{%
}{.}\hspace{.4pt}1109\discretionary{/}{%
}{/}TVCG\hspace{.1pt}\discretionary{.}{%
}{.}\hspace{.4pt}2022\hspace{.1pt}\discretionary{.}{%
}{.}\hspace{.4pt}3207157}}


\bibitem{Alaee2018AUS}
G.~Alaee, A.~P. Deasi, L.~Peña-Castillo, E.~Brown, and O.~Meruvia-Pastor.
\newblock A user study on augmented virtuality using depth sensing cameras for near-range awareness in immersive vr.
\newblock 2018.

\bibitem{Batschelet1981CircularSI}
E.~Batschelet.
\newblock {\em Circular statistics in biology}.
\newblock 1981. doi: {{%
10\hspace{.1pt}\discretionary{.}{%
}{.}\hspace{.4pt}2307\discretionary{/}{%
}{/}2981498}}


\bibitem{Bimberg2021VirtualRF}
P.~Bimberg, T.~Weissker, A.~Kulik, and B.~Froehlich.
\newblock Virtual rotations for maneuvering in immersive virtual environments.
\newblock {\em Proceedings of the 27th ACM Symposium on Virtual Reality Software and Technology}, 2021. doi: {{%
10\hspace{.1pt}\discretionary{.}{%
}{.}\hspace{.4pt}1145\discretionary{/}{%
}{/}3489849\hspace{.1pt}\discretionary{.}{%
}{.}\hspace{.4pt}3489893}}


\bibitem{Bozgeyikli2016PointT}
E.~Bozgeyikli, A.~Raij, S.~Katkoori, and R.~Dubey.
\newblock Point \& teleport locomotion technique for virtual reality.
\newblock {\em Proceedings of the 2016 Annual Symposium on Computer-Human Interaction in Play}, 2016. doi: {{%
10\hspace{.1pt}\discretionary{.}{%
}{.}\hspace{.4pt}1145\discretionary{/}{%
}{/}2967934\hspace{.1pt}\discretionary{.}{%
}{.}\hspace{.4pt}2968105}}


\bibitem{Bruder2015CognitiveRD}
G.~Bruder, P.~Lubos, and F.~Steinicke.
\newblock Cognitive resource demands of redirected walking.
\newblock {\em IEEE Transactions on Visualization and Computer Graphics}, 21:539--544, 2015. doi: {{%
10\hspace{.1pt}\discretionary{.}{%
}{.}\hspace{.4pt}1109\discretionary{/}{%
}{/}TVCG\hspace{.1pt}\discretionary{.}{%
}{.}\hspace{.4pt}2015\hspace{.1pt}\discretionary{.}{%
}{.}\hspace{.4pt}2391864}}


\bibitem{Budhiraja2015WheresMD}
P.~Budhiraja, R.~Sodhi, B.~R. Jones, K.~Karsch, B.~Bailey, and D.~Forsyth.
\newblock Where's my drink? enabling peripheral real world interactions while using hmds.
\newblock {\em ArXiv}, abs/1502.04744, 2015.

\bibitem{Cha2022DesignAU}
B.~Cha, Y.~Bae, C.-G. Lee, D.~Jeong, and J.~Ryu.
\newblock Design and user evaluation of haptic augmented virtuality system for immersive virtual training.
\newblock {\em International Journal of Control, Automation and Systems}, 20:3032 -- 3044, 2022. doi: {{%
10\hspace{.1pt}\discretionary{.}{%
}{.}\hspace{.4pt}1007\discretionary{/}{%
}{/}s12555\discretionary{%
}{-}{-}021\discretionary{%
}{-}{-}0072\discretionary{%
}{-}{-}6}}


\bibitem{Chiossi2024EvaluatingTP}
F.~Chiossi, Y.~E. Khaoudi, C.~Ou, L.~Sidenmark, A.~Zaky, T.~M. Feuchtner, and S.~Mayer.
\newblock Evaluating typing performance in different mixed reality manifestations using physiological features.
\newblock {\em Proc. ACM Hum. Comput. Interact.}, 8:377--406, 2024. doi: {{%
10\hspace{.1pt}\discretionary{.}{%
}{.}\hspace{.4pt}1145\discretionary{/}{%
}{/}3698142}}


\bibitem{Desai2017AWT}
A.~P. Desai, L.~P. Castillo, and O.~E.~M. Pastor.
\newblock A window to your smartphone: Exploring interaction and communication in immersive vr with augmented virtuality.
\newblock {\em 2017 14th Conference on Computer and Robot Vision (CRV)}, pp. 217--224, 2017. doi: {{%
10\hspace{.1pt}\discretionary{.}{%
}{.}\hspace{.4pt}1109\discretionary{/}{%
}{/}CRV\hspace{.1pt}\discretionary{.}{%
}{.}\hspace{.4pt}2017\hspace{.1pt}\discretionary{.}{%
}{.}\hspace{.4pt}16}}


\bibitem{Eichhorn2023ShoppingIB}
C.~Eichhorn, D.~A. Plecher, T.~Mesmer, L.~Leder, T.~Simecek, N.~Boukadida, and G.~Klinker.
\newblock Shopping in between realities-using an augmented virtuality smartphone in a virtual supermarket.
\newblock {\em 2023 IEEE International Symposium on Mixed and Augmented Reality (ISMAR)}, pp. 1161--1170, 2023. doi: {{%
10\hspace{.1pt}\discretionary{.}{%
}{.}\hspace{.4pt}1109\discretionary{/}{%
}{/}ISMAR59233\hspace{.1pt}\discretionary{.}{%
}{.}\hspace{.4pt}2023\hspace{.1pt}\discretionary{.}{%
}{.}\hspace{.4pt}00133}}


\bibitem{Feld2024EffectsOH}
N.~Feld, D.~Zielasko, and B.~Weyers.
\newblock Effects of hand occlusion in radial mid-air menu interaction in augmented reality.
\newblock {\em 2024 IEEE Conference on Virtual Reality and 3D User Interfaces Abstracts and Workshops (VRW)}, pp. 551--558, 2024. doi: {{%
10\hspace{.1pt}\discretionary{.}{%
}{.}\hspace{.4pt}1109\discretionary{/}{%
}{/}VRW62533\hspace{.1pt}\discretionary{.}{%
}{.}\hspace{.4pt}2024\hspace{.1pt}\discretionary{.}{%
}{.}\hspace{.4pt}00106}}


\bibitem{hart1988development}
S.~G. Hart and L.~E. Staveland.
\newblock Development of nasa-tlx (task load index): Results of empirical and theoretical research.
\newblock {\em Advances in Psychology}, 52:139--183, 1988. doi: {{%
10\hspace{.1pt}\discretionary{.}{%
}{.}\hspace{.4pt}1016\discretionary{/}{%
}{/}S0166\discretionary{%
}{-}{-}4115\discretionary{%
}{(}{(}08\discretionary{)}{%
}{)}62386\discretionary{%
}{-}{-}9}}


\bibitem{Hoffman1998PhysicallyTV}
H.~Hoffman.
\newblock Physically touching virtual objects using tactile augmentation enhances the realism of virtual environments.
\newblock {\em Proceedings. IEEE 1998 Virtual Reality Annual International Symposium (Cat. No.98CB36180)}, pp. 59--63, 1998. doi: {{%
10\hspace{.1pt}\discretionary{.}{%
}{.}\hspace{.4pt}1109\discretionary{/}{%
}{/}VRAIS\hspace{.1pt}\discretionary{.}{%
}{.}\hspace{.4pt}1998\hspace{.1pt}\discretionary{.}{%
}{.}\hspace{.4pt}658423}}


\bibitem{Insko2001PassiveHS}
B.~Insko, M.~Meehan, M.~Whitton, and F.~Brooks.
\newblock {\em Passive haptics significantly enhances virtual environments}.
\newblock 2001.

\bibitem{Kanamori2018ObstacleAM}
K.~Kanamori, N.~Sakata, T.~Tominaga, Y.~Hijikata, K.~Harada, and K.~Kiyokawa.
\newblock Obstacle avoidance method in real space for virtual reality immersion.
\newblock {\em 2018 IEEE International Symposium on Mixed and Augmented Reality (ISMAR)}, pp. 80--89, 2018. doi: {{%
10\hspace{.1pt}\discretionary{.}{%
}{.}\hspace{.4pt}1109\discretionary{/}{%
}{/}ISMAR\hspace{.1pt}\discretionary{.}{%
}{.}\hspace{.4pt}2018\hspace{.1pt}\discretionary{.}{%
}{.}\hspace{.4pt}00033}}


\bibitem{Kang2020SafeXRAW}
H.~Kang and J.~Han.
\newblock Safexr: alerting walking persons to obstacles in mobile xr environments.
\newblock {\em The Visual Computer}, pp. 1--13, 2020. doi: {{%
10\hspace{.1pt}\discretionary{.}{%
}{.}\hspace{.4pt}1007\discretionary{/}{%
}{/}s00371\discretionary{%
}{-}{-}020\discretionary{%
}{-}{-}01907\discretionary{%
}{-}{-}4}}


\bibitem{fms}
B.~Keshavarz and H.~Hecht.
\newblock Validating an efficient method to quantify motion sickness.
\newblock {\em Human Factors}, 53:415--426, 2011. doi: {{%
10\hspace{.1pt}\discretionary{.}{%
}{.}\hspace{.4pt}1177\discretionary{/}{%
}{/}0018720811403736}}


\bibitem{Kohli2005CombiningPH}
L.~Kohli, E.~Burns, D.~Miller, and H.~Fuchs.
\newblock Combining passive haptics with redirected walking.
\newblock pp. 253--254, 2005. doi: {{%
10\hspace{.1pt}\discretionary{.}{%
}{.}\hspace{.4pt}1145\discretionary{/}{%
}{/}1152399\hspace{.1pt}\discretionary{.}{%
}{.}\hspace{.4pt}1152451}}


\bibitem{Langbehn2018EvaluationOL}
E.~Langbehn, P.~Lubos, and F.~Steinicke.
\newblock Evaluation of locomotion techniques for room-scale vr: Joystick, teleportation, and redirected walking.
\newblock {\em Proceedings of the Virtual Reality International Conference - Laval Virtual}, 2018. doi: {{%
10\hspace{.1pt}\discretionary{.}{%
}{.}\hspace{.4pt}1145\discretionary{/}{%
}{/}3234253\hspace{.1pt}\discretionary{.}{%
}{.}\hspace{.4pt}3234291}}


\bibitem{Lewis2013UMUXLITEWT}
J.~R. Lewis, B.~Utesch, and D.~E. Maher.
\newblock Umux-lite: when there's no time for the sus.
\newblock In {\em International Conference on Human Factors in Computing Systems}, 2013. doi: {{%
10\hspace{.1pt}\discretionary{.}{%
}{.}\hspace{.4pt}1145\discretionary{/}{%
}{/}2470654\hspace{.1pt}\discretionary{.}{%
}{.}\hspace{.4pt}2481287}}


\bibitem{mcgill2015dose}
M.~McGill, D.~Boland, R.~Murray-Smith, and S.~Brewster.
\newblock A dose of reality.
\newblock In {\em Proceedings of the 33rd Annual ACM Conference on Human Factors in Computing Systems - CHI '15}, pp. 2143--2152. ACM Press, Seoul, Republic of Korea, 2015.
\newblock Conference date: 2015.04.18-2015.04.23. doi: {{%
10\hspace{.1pt}\discretionary{.}{%
}{.}\hspace{.4pt}1145\discretionary{/}{%
}{/}2702123\hspace{.1pt}\discretionary{.}{%
}{.}\hspace{.4pt}2702382}}


\bibitem{Milgram1995AugmentedRA}
P.~Milgram, H.~Takemura, A.~Utsumi, and F.~Kishino.
\newblock Augmented reality: a class of displays on the reality-virtuality continuum.
\newblock In {\em Other Conferences}, vol. 2351, 1995. doi: {{%
10\hspace{.1pt}\discretionary{.}{%
}{.}\hspace{.4pt}1117\discretionary{/}{%
}{/}12\hspace{.1pt}\discretionary{.}{%
}{.}\hspace{.4pt}197321}}


\bibitem{Nakano2022UkemochiAV}
K.~Nakano, D.~Horita, N.~Isoyama, H.~Uchiyama, and K.~Kiyokawa.
\newblock Ukemochi: A video see-through food overlay system for eating experience in the metaverse.
\newblock {\em CHI Conference on Human Factors in Computing Systems Extended Abstracts}, 2022. doi: {{%
10\hspace{.1pt}\discretionary{.}{%
}{.}\hspace{.4pt}1145\discretionary{/}{%
}{/}3491101\hspace{.1pt}\discretionary{.}{%
}{.}\hspace{.4pt}3519779}}


\bibitem{Nilsson201815YO}
N.~C. Nilsson, T.~C. Peck, G.~Bruder, E.~Hodgson, S.~Serafin, M.~Whitton, F.~Steinicke, and E.~S. Rosenberg.
\newblock 15 years of research on redirected walking in immersive virtual environments.
\newblock {\em IEEE Computer Graphics and Applications}, 38:44--56, 2018. doi: {{%
10\hspace{.1pt}\discretionary{.}{%
}{.}\hspace{.4pt}1109\discretionary{/}{%
}{/}MCG\hspace{.1pt}\discretionary{.}{%
}{.}\hspace{.4pt}2018\hspace{.1pt}\discretionary{.}{%
}{.}\hspace{.4pt}111125628}}


\bibitem{Palma2021AugmentedVU}
G.~Palma, S.~Perry, and P.~Cignoni.
\newblock Augmented virtuality using touch-sensitive 3d-printed objects.
\newblock {\em Remote. Sens.}, 13:2186, 2021. doi: {{%
10\hspace{.1pt}\discretionary{.}{%
}{.}\hspace{.4pt}3390\discretionary{/}{%
}{/}rs13112186}}


\bibitem{Pointecker2023VisualMF}
F.~Pointecker, D.~Oberögger, and C.~Anthes.
\newblock Visual metaphors for notification into virtual environments.
\newblock {\em 2023 IEEE International Symposium on Mixed and Augmented Reality Adjunct (ISMAR-Adjunct)}, pp. 60--64, 2023. doi: {{%
10\hspace{.1pt}\discretionary{.}{%
}{.}\hspace{.4pt}1109\discretionary{/}{%
}{/}ISMAR\discretionary{%
}{-}{-}Adjunct60411\hspace{.1pt}\discretionary{.}{%
}{.}\hspace{.4pt}2023\hspace{.1pt}\discretionary{.}{%
}{.}\hspace{.4pt}00020}}


\bibitem{prithul2021teleportation}
A.~Prithul, I.~B. Adhanom, and E.~Folmer.
\newblock {Teleportation in Virtual Reality; A Mini-Review}.
\newblock {\em Frontiers in Virtual Reality}, 2:730792, 2021.

\bibitem{Rahimi2020SceneTA}
K.~Rahimi, C.~Banigan, and E.~Ragan.
\newblock Scene transitions and teleportation in virtual reality and the implications for spatial awareness and sickness.
\newblock {\em IEEE Transactions on Visualization and Computer Graphics}, 26:2273--2287, 2020. doi: {{%
10\hspace{.1pt}\discretionary{.}{%
}{.}\hspace{.4pt}1109\discretionary{/}{%
}{/}TVCG\hspace{.1pt}\discretionary{.}{%
}{.}\hspace{.4pt}2018\hspace{.1pt}\discretionary{.}{%
}{.}\hspace{.4pt}2884468}}


\bibitem{Roo2017OneRA}
J.~Roo and M.~Hachet.
\newblock {\em One Reality: Augmenting How the Physical World is Experienced by combining Multiple Mixed Reality Modalities}.
\newblock 2017. doi: {{%
10\hspace{.1pt}\discretionary{.}{%
}{.}\hspace{.4pt}1145\discretionary{/}{%
}{/}3126594\hspace{.1pt}\discretionary{.}{%
}{.}\hspace{.4pt}3126638}}


\bibitem{Shin2022IncorporatingRO}
J.~Shin and K.~Lee.
\newblock Incorporating real-world object into virtual reality: using mobile device input with augmented virtuality.
\newblock {\em Multimedia Tools and Applications}, 83:46625--46652, 2022. doi: {{%
10\hspace{.1pt}\discretionary{.}{%
}{.}\hspace{.4pt}1007\discretionary{/}{%
}{/}s11042\discretionary{%
}{-}{-}022\discretionary{%
}{-}{-}13637\discretionary{%
}{-}{-}x}}


\bibitem{Simeone2015substitutionalReality}
A.~L. Simeone, E.~Velloso, and H.~Gellersen.
\newblock Substitutional reality: Using the physical environment to design virtual reality experiences.
\newblock In {\em Proceedings of ACM Conference on Human Factors in Computing Systems}, p. 3307–3316. Association for Computing Machinery, New York, NY, USA, 2015. doi: {{%
10\hspace{.1pt}\discretionary{.}{%
}{.}\hspace{.4pt}1145\discretionary{/}{%
}{/}2702123\hspace{.1pt}\discretionary{.}{%
}{.}\hspace{.4pt}2702389}}


\bibitem{Slater2004HowCW}
M.~Slater.
\newblock How colorful was your day? why questionnaires cannot assess presence in virtual environments.
\newblock {\em Presence: Teleoperators \& Virtual Environments}, 13:484--493, 2004. doi: {{%
10\hspace{.1pt}\discretionary{.}{%
}{.}\hspace{.4pt}1162\discretionary{/}{%
}{/}1054746041944849}}


\bibitem{Sousa2019SafeWI}
M.~Sousa, D.~Mendes, and J.~Jorge.
\newblock Safe walking in vr using augmented virtuality.
\newblock {\em ArXiv}, abs/1911.13032, 2019.

\bibitem{Steinicke2010EstimationOD}
F.~Steinicke, G.~Bruder, J.~Jerald, H.~Frenz, and M.~Lappe.
\newblock Estimation of detection thresholds for redirected walking techniques.
\newblock {\em IEEE Transactions on Visualization and Computer Graphics}, 16:17--27, 2010. doi: {{%
10\hspace{.1pt}\discretionary{.}{%
}{.}\hspace{.4pt}1109\discretionary{/}{%
}{/}TVCG\hspace{.1pt}\discretionary{.}{%
}{.}\hspace{.4pt}2009\hspace{.1pt}\discretionary{.}{%
}{.}\hspace{.4pt}62}}


\bibitem{Tapal2017TheSO}
A.~Tapal, E.~Oren, R.~Dar, and B.~Eitam.
\newblock The sense of agency scale: A measure of consciously perceived control over one's mind, body, and the immediate environment.
\newblock {\em Frontiers in Psychology}, 8, 2017. doi: {{%
10\hspace{.1pt}\discretionary{.}{%
}{.}\hspace{.4pt}3389\discretionary{/}{%
}{/}fpsyg\hspace{.1pt}\discretionary{.}{%
}{.}\hspace{.4pt}2017\hspace{.1pt}\discretionary{.}{%
}{.}\hspace{.4pt}01552}}


\bibitem{Thomas2020TowardsPI}
J.~Thomas, C.~H. Pospick, and E.~S. Rosenberg.
\newblock Towards physically interactive virtual environments: Reactive alignment with redirected walking.
\newblock {\em Proceedings of the 26th ACM Symposium on Virtual Reality Software and Technology}, 2020. doi: {{%
10\hspace{.1pt}\discretionary{.}{%
}{.}\hspace{.4pt}1145\discretionary{/}{%
}{/}3385956\hspace{.1pt}\discretionary{.}{%
}{.}\hspace{.4pt}3418966}}


\bibitem{Thorndike1953WhoBI}
R.~L. Thorndike.
\newblock Who belongs in the family?
\newblock {\em Psychometrika}, 18:267--276, 1953. doi: {{%
10\hspace{.1pt}\discretionary{.}{%
}{.}\hspace{.4pt}1007\discretionary{/}{%
}{/}BF02289263}}


\bibitem{Tian2019EnhancingAV}
Y.~Tian, C.-W. Fu, S.~Zhao, R.~Li, X.~Tang, X.~Hu, and P.~Heng.
\newblock Enhancing augmented vr interaction via egocentric scene analysis.
\newblock {\em Proceedings of the ACM on Interactive, Mobile, Wearable and Ubiquitous Technologies}, 3:1 -- 24, 2019. doi: {{%
10\hspace{.1pt}\discretionary{.}{%
}{.}\hspace{.4pt}1145\discretionary{/}{%
}{/}3351263}}


\bibitem{Willich2019YouIM}
J.~von Willich, M.~Funk, F.~Müller, K.~Marky, J.~Riemann, and M.~Mühlhäuser.
\newblock You invaded my tracking space! using augmented virtuality for spotting passersby in room-scale virtual reality.
\newblock In {\em Conference on Designing Interactive Systems}, 2019. doi: {{%
10\hspace{.1pt}\discretionary{.}{%
}{.}\hspace{.4pt}1145\discretionary{/}{%
}{/}3322276\hspace{.1pt}\discretionary{.}{%
}{.}\hspace{.4pt}3322334}}


\bibitem{Wang2022RealityLensAU}
C.-H. Wang, B.-Y. Chen, and L.~Chan.
\newblock Realitylens: A user interface for blending customized physical world view into virtual reality.
\newblock In {\em ACM Symposium on User Interface Software and Technology}, 2022. doi: {{%
10\hspace{.1pt}\discretionary{.}{%
}{.}\hspace{.4pt}1145\discretionary{/}{%
}{/}3526113\hspace{.1pt}\discretionary{.}{%
}{.}\hspace{.4pt}3545686}}


\bibitem{Weissker2018SpatialUA}
T.~Weissker, A.~Kunert, B.~Fröhlich, and A.~Kulik.
\newblock Spatial updating and simulator sickness during steering and jumping in immersive virtual environments.
\newblock {\em 2018 IEEE Conference on Virtual Reality and 3D User Interfaces (VR)}, pp. 97--104, 2018. doi: {{%
10\hspace{.1pt}\discretionary{.}{%
}{.}\hspace{.4pt}1109\discretionary{/}{%
}{/}VR\hspace{.1pt}\discretionary{.}{%
}{.}\hspace{.4pt}2018\hspace{.1pt}\discretionary{.}{%
}{.}\hspace{.4pt}8446620}}


\bibitem{Wheeler2024ProporientedWR}
S.~G. Wheeler, S.~Hoermann, R.~W. Lindeman, G.~Ghinea, and A.~Covaci.
\newblock Prop-oriented world rotation: enabling passive haptic feedback by aligning real and virtual objects in virtual reality.
\newblock {\em Multimedia Tools and Applications}, 2024. doi: {{%
10\hspace{.1pt}\discretionary{.}{%
}{.}\hspace{.4pt}1007\discretionary{/}{%
}{/}s11042\discretionary{%
}{-}{-}024\discretionary{%
}{-}{-}18200\discretionary{%
}{-}{-}4}}


\bibitem{Williams2021ARCAR}
N.~L. Williams, A.~Bera, and D.~Manocha.
\newblock Arc: Alignment-based redirection controller for redirected walking in complex environments.
\newblock {\em IEEE Transactions on Visualization and Computer Graphics}, 27:2535--2544, 2021. doi: {{%
10\hspace{.1pt}\discretionary{.}{%
}{.}\hspace{.4pt}1109\discretionary{/}{%
}{/}TVCG\hspace{.1pt}\discretionary{.}{%
}{.}\hspace{.4pt}2021\hspace{.1pt}\discretionary{.}{%
}{.}\hspace{.4pt}3067781}}


\bibitem{Wozniak2018TowardsUO}
P.~Wozniak, A.~Capobianco, N.~Javahiraly, and D.~Curticapean.
\newblock Towards unobtrusive obstacle detection and notification for vr.
\newblock {\em Proceedings of the 24th ACM Symposium on Virtual Reality Software and Technology}, 2018. doi: {{%
10\hspace{.1pt}\discretionary{.}{%
}{.}\hspace{.4pt}1145\discretionary{/}{%
}{/}3281505\hspace{.1pt}\discretionary{.}{%
}{.}\hspace{.4pt}3283391}}


\bibitem{Yechiam2014TheCB}
E.~Yechiam, A.~Telpaz, and G.~Hochman.
\newblock The complaint bias in subjective evaluations of incentives.
\newblock vol.~1, pp. 147--160, 2014. doi: {{%
10\hspace{.1pt}\discretionary{.}{%
}{.}\hspace{.4pt}1037\discretionary{/}{%
}{/}DEC0000008}}


\bibitem{Zhang2021InTW}
Y.~Zhang, S.-T. Ho, N.~Ladevèze, H.~Nguyen, C.~Fleury, and P.~Bourdot.
\newblock In touch with everyday objects: Teleportation techniques in virtual environments supporting tangibility.
\newblock {\em 2021 IEEE Conference on Virtual Reality and 3D User Interfaces Abstracts and Workshops (VRW)}, pp. 278--283, 2021. doi: {{%
10\hspace{.1pt}\discretionary{.}{%
}{.}\hspace{.4pt}1109\discretionary{/}{%
}{/}VRW52623\hspace{.1pt}\discretionary{.}{%
}{.}\hspace{.4pt}2021\hspace{.1pt}\discretionary{.}{%
}{.}\hspace{.4pt}00057}}


\bibitem{Zhang2020VirtualNC}
Y.~Zhang, N.~Ladevèze, H.~Nguyen, C.~Fleury, and P.~Bourdot.
\newblock Virtual navigation considering user workspace: Automatic and manual positioning before teleportation.
\newblock {\em Proceedings of the 26th ACM Symposium on Virtual Reality Software and Technology}, 2020. doi: {{%
10\hspace{.1pt}\discretionary{.}{%
}{.}\hspace{.4pt}1145\discretionary{/}{%
}{/}3385956\hspace{.1pt}\discretionary{.}{%
}{.}\hspace{.4pt}3418949}}


\bibitem{Zielasko2022SystematicDS}
D.~Zielasko, J.~Heib, and B.~Weyers.
\newblock Systematic design space exploration of discrete virtual rotations in vr.
\newblock {\em 2022 IEEE Conference on Virtual Reality and 3D User Interfaces (VR)}, pp. 693--702, 2022. doi: {{%
10\hspace{.1pt}\discretionary{.}{%
}{.}\hspace{.4pt}1109\discretionary{/}{%
}{/}VR51125\hspace{.1pt}\discretionary{.}{%
}{.}\hspace{.4pt}2022\hspace{.1pt}\discretionary{.}{%
}{.}\hspace{.4pt}00090}}


\bibitem{zielasko2019menu}
D.~Zielasko, M.~Kr{\"u}ger, B.~Weyers, and T.~W. Kuhlen.
\newblock {Passive Haptic Menus for Desk-Based and HMD-Projected Virtual Reality}.
\newblock {\em Proc. of IEEE VR Workshop on Everyday Virtual Reality}, pp. 1--6, 2019. doi: {{%
10\hspace{.1pt}\discretionary{.}{%
}{.}\hspace{.4pt}1109\discretionary{/}{%
}{/}WEVR\hspace{.1pt}\discretionary{.}{%
}{.}\hspace{.4pt}2019\hspace{.1pt}\discretionary{.}{%
}{.}\hspace{.4pt}8809589}}


\bibitem{Zielasko2024CarryOverER}
D.~Zielasko, B.~Rehling, D.~Clement, and G.~Domes.
\newblock Carry-over effects ruin your (cybersickness) experiments and balancing conditions is not a solution.
\newblock In {\em 2024 IEEE Conference on Virtual Reality and 3D User Interfaces Abstracts and Workshops (VRW)}, pp. 1--5, 2024. doi: {{%
10\hspace{.1pt}\discretionary{.}{%
}{.}\hspace{.4pt}1109\discretionary{/}{%
}{/}VRW62533\hspace{.1pt}\discretionary{.}{%
}{.}\hspace{.4pt}2024\hspace{.1pt}\discretionary{.}{%
}{.}\hspace{.4pt}00007}}


\bibitem{Zielasko2024DiscreteVR}
D.~Zielasko, M.~Späth, and M.~Wölwer.
\newblock Discrete virtual rotation in pointing vs. leaning-directed steering interfaces: A uni vs. bimanual perspective.
\newblock {\em ArXiv}, abs/2406.14212, 2024. doi: {{%
10\hspace{.1pt}\discretionary{.}{%
}{.}\hspace{.4pt}48550\discretionary{/}{%
}{/}arXiv\hspace{.1pt}\discretionary{.}{%
}{.}\hspace{.4pt}2406\hspace{.1pt}\discretionary{.}{%
}{.}\hspace{.4pt}14212}}


\bibitem{zielasko2017}
D.~Zielasko, B.~Weyers, M.~Bellgardt, S.~Pick, A.~Mei{\ss}ner, T.~Vierjahn, and T.~W. Kuhlen.
\newblock {Remain Seated: Towards Fully-Immersive Desktop VR}.
\newblock {\em Proc. of IEEE VR Workshop on Everyday Virtual Reality}, pp. 1--6, 2017. doi: {{%
10\hspace{.1pt}\discretionary{.}{%
}{.}\hspace{.4pt}1109\discretionary{/}{%
}{/}WEVR\hspace{.1pt}\discretionary{.}{%
}{.}\hspace{.4pt}2017\hspace{.1pt}\discretionary{.}{%
}{.}\hspace{.4pt}7957707}}


\end{thebibliography}
\end{document}